\documentclass[times,twocolumn]{aastex63}
\hypersetup{linkcolor=red,citecolor=blue,filecolor=cyan,urlcolor=blue}
\usepackage{amsmath}
\usepackage{graphicx}
\usepackage{txfonts}
\usepackage{amssymb}
\usepackage[]{natbib}
\usepackage{mathrsfs}
\usepackage{float}
\usepackage{amstext}

\makeatletter
\newcommand\footnoteref[1]{\protected@xdef\@thefnmark{\ref{#1}}\@footnotemark}
\makeatother

\usepackage{xspace}

\newcommand{\fref}[1]{Figure~\ref{#1}}

\newcommand{\sref}[1]{Section~\ref{#1}}

\newcommand{\myrev}[1]{{\bf}}
\received{April 20, 2020}
\accepted{\today}

\submitjournal{AJ}
\shorttitle{Timing verification of TESS}
\shortauthors{von Essen et al.}

\begin{document}

\title{TESS Data for Asteroseismology: Timing verification\footnote{Based on observations made at the Argentinian Complejo Astron\'omico El Leoncito (CASLEO)}}

\correspondingauthor{Carolina von Essen}
\email{cessen@phys.au.dk}

\author[0000-0002-6956-1725]{Carolina von Essen}
\affil{Stellar Astrophysics Centre, Department of Physics and Astronomy, Aarhus University, Ny Munkegade 120, DK-8000 Aarhus C, Denmark.}
\affil{Astronomical Observatory, Institute of Theoretical Physics and Astronomy, Vilnius University, Sauletekio av. 3, 10257, Vilnius, Lithuania}

\author[0000-0001-9214-5642]{Mikkel N. Lund}
\affil{Stellar Astrophysics Centre, Department of Physics and Astronomy, Aarhus University, Ny Munkegade 120, DK-8000 Aarhus C, Denmark.}

\author[0000-0001-8725-4502]{Rasmus Handberg}
\affil{Stellar Astrophysics Centre, Department of Physics and Astronomy, Aarhus University, Ny Munkegade 120, DK-8000 Aarhus C, Denmark.}

\author[0000-0001-5201-5472]{Marina S. Sosa}
\affil{Stellar Astrophysics Centre, Department of Physics and Astronomy, Aarhus University, Ny Munkegade 120, DK-8000 Aarhus C, Denmark.}

\author{Julie Thiim Gadeberg}
\affil{Stellar Astrophysics Centre, Department of Physics and Astronomy, Aarhus University, Ny Munkegade 120, DK-8000 Aarhus C, Denmark.}

\author[0000-0002-9037-0018]{Hans Kjeldsen}
\affil{Stellar Astrophysics Centre, Department of Physics and Astronomy, Aarhus University, Ny Munkegade 120, DK-8000 Aarhus C, Denmark.}
\affil{Astronomical Observatory, Institute of Theoretical Physics and Astronomy, Vilnius University, Sauletekio av. 3, 10257, Vilnius, Lithuania}

\author[0000-0001-6763-6562]{Roland K. Vanderspek}
\affil{MIT Kavli Institute for Astrophysics and Space Research, Massachusetts Institute of Technology, 77 Massachusetts Avenue, 37-241 Cambridge, MA 02139}

\author{Dina S. Mortensen}
\affil{Stellar Astrophysics Centre, Department of Physics and Astronomy, Aarhus University, Ny Munkegade 120, DK-8000 Aarhus C, Denmark.}

\author[0000-0003-2865-042X]{M. Mallonn}
\affil{Leibniz-Institut f\"{u}r Astrophysik Potsdam (AIP), An der Sternwarte 16, D-14482 Potsdam, Germany.}

\author{L. Mammana}
\affil{Complejo Astron\'omico El Leoncito (CONICET-UNLP-UNC-UNSJ), Av. Espa\~na 1512 Sur, San Juan, Argentina.}
\affil{Facultad de Ciencias Astron\'omicas y Geof\'isicas, Universidad Nacional de La Plata, Paseo del Bosque, B1900FWA, La Plata, Argentina}

\author{Edward H. Morgan}
\author{Jesus Noel S. Villase\~{n}or}
\author[0000-0002-9113-7162]{Michael M. Fausnaugh}
\author{George R. Ricker}
\affil{MIT Kavli Institute for Astrophysics and Space Research, Massachusetts Institute of Technology, 77 Massachusetts Avenue, 37-241 Cambridge, MA 02139}

\begin{abstract}

The Transiting Exoplanet Survey Satellite (TESS) is NASA's latest space telescope dedicated to the discovery of transiting exoplanets around nearby stars. Besides the main goal of the mission, asteroseismology is an important secondary goal and very relevant for the high-quality time series that TESS will make during its two year all-sky survey. Using TESS for asteroseismology introduces strong timing requirements, especially for coherent oscillators. Although the internal clock on board TESS is precise in its own time, it might have a constant drift and will thus need calibration, or offsets might inadvertently be introduced. Here we present simultaneously ground- and space-based observations of primary eclipses of several binary systems in the Southern ecliptic hemisphere, used to verify the reliability of the TESS timestamps. From twelve contemporaneous TESS/ground observations we determined a time offset equal to 5.8 $\pm$ 2.5 sec, in the sense that the Barycentric time measured by TESS is ahead of real time. The offset is consistent with zero at $2.3-\sigma$ level. In addition, we used 405 individually measured mid-eclipse times of 26 eclipsing binary stars observed solely by TESS to test the existence of a potential drift with a monotonic growth (or decay) affecting the observations of all stars. We find a drift corresponding to $\rm \sigma_{\rm drift}=0.009\pm 0.015\,\, sec/day$. We find that the measured offset is of a size that will not become an issue for comparing ground-based and space data for coherent oscillations for most of the targets observed with TESS.
\end{abstract}

\keywords{stars: binary systems -- stars: individual: BD Dor, KX Aqr, NV Tel, WY Cet, VV Eri, AO Pic, AW Vel, X Pic, V636 Cen, RR Nor, TV Nor -- methods: observational}

\section{Introduction}

Owing to the high precision and long duration time series provided during the last decade by space missions such as {\it Kepler} \citep{Borucki2010,Koch2010}, K2 \citep{2014PASP..126..398H}, and CoRoT \citep{Auvergne2009}, the field of asteroseismology has led a revolution in stellar astrophysics. The power of the method relies in accessing the stellar interiors through the study of the surface manifestation of internal resonant oscillations. In addition to its contribution to stellar physics \citep[e.g.,][]{chaplin2013,Hekker2017,Bowman2017,Garcia2019}, asteroseismology has also helped advance the field of exoplanets \citep{VanEylen2014,Lundkvist2016}. 

In April, 2018, the Transiting Exoplanet Survey Satellite (TESS) joined the short list of space-based telescopes dedicated to finding planets by means of the transit method \citep{Ricker2015}.  TESS hosts four charge coupled device (CCD) cameras aligned with the ecliptical poles, that stare at the same fraction of the sky for two of TESS orbits (2$\times$13.7 days, approximately). The observations collected by the four CCDs during two consecutive orbits are defined as a sector. Due to the large field of view of the CCDs (24$\times$24 degrees each), the Ecliptic hemispheres are divided in 13 sectors, specifically 13 in the southern hemisphere and 13 in the northern hemisphere during the primary mission. Different from {\it Kepler}, TESS is  designed to detect transiting planets around very bright stars, which permits us to easily carry out ground-based radial velocity follow-ups to determine planetary masses \citep{Trifonov2019,Rodriguez2019}. However, using TESS for asteroseismology introduces strong timing requirements \citep{Lund2017}. Although the internal clock of TESS might be very accurate in its own time, it can have a constant drift or offset or variation in the length of a second, caused by hardware limitations, software errors, lags in electronics after safe-modes/downlinks,  missed leap seconds, and wrong reference frames, among others. In consequence, time stamps need verification and possibly calibration.

The TESS Asteroseismic Science Consortium (TASC) hosts the group ``TESS Data for Asteroseismology'' (T'DA) which is in charge of delivering light curves for all of TASC, hence encompassing many different types of stars, including all targets found in full frame images. Requested by the TESS Science Processing Operations Center (SPOC), T'DA was also asked to carry out independent verification of TESS timestamps. This exercise is required to ensure the highest level of asteroseismic inference from TESS data, and works as a mechanism to prevent and diagnose any timing malfunction, as it happened to {\it Kepler} timestamps\footnote{\url{https://archive.stsci.edu/kepler/release_notes/release_notes19/DataRelease_19_20130204.pdf}}. To carry out this work, TESS has been continuously observing a modest list of eclipsing binary systems (EBSs) with relatively short periods, most of them between 0.7 and 4.5 days with the exception of TV Nor, which has an orbital period of 8.5 days. In order to achieve accurate timing measurements, the EBSs are mostly of Algol type presenting deep, V-shaped, and relatively short eclipses. They cover a range of latitudinal and longitudinal ecliptic coordinates, to ensure observability throughout TESS's first year. 

In this work we present the timing requirements to be able to carry out asteroseismology using TESS data in Section~\ref{sec:req}, and we show the photometric data collected from two ground-based telescopes located in Argentina and gathered by TESS in Section~\ref{sec:Data}. We detail our strategy for determining the mid-eclipse times and the model functions used in Section~\ref{sec:Mid-Eclipses}, and we present the timing verification computed from contemporaneous ground and ground, and ground and space-based data in Sections~\ref{sec:time_test} and~\ref{sec:time_offset}, along with the timing verification carried out solely using TESS data in Section~\ref{sec:time_drift}. We close this work with our final remarks in Section~\ref{sec:DyC}.

\section{Timing requirements for asteroseismology}
\label{sec:req}

The requirements for timing for asteroseismology are mainly of importance for high amplitude coherent oscillators (such as $\delta$ Sct and RR-lyr stars), while the requirements for stochastic oscillators are less strict. The formal requirements are specified in the internal document SAC\_TESS\_0002\_5\footnote{\url{https://tasoc.dk/docs/SAC_TESS_0002_5.pdf}}, which discuss three main categories: (1) accurate values for the exposure length, required to reach the photon noise limit (requirement RS-TASC-01); (2) accurate knowledge of differential times within one month of observations, required to reach the theoretical accuracy on oscillation mode frequencies and amplitudes (especially important for coherent oscillators). Important here is also the conversion of spacecraft times to barycentric julian date (BJD), which should be as accurate as the determination of differential times (requirements RS-TASC-02 and RS-TASC-03)\footnote{Measured to 10 ms (rms) (R.~K.~Vanderspek, private communication)}; (3) to compare observations from TESS with ground-based facilities the absolute time in BJD needed (requirement RS-TASC-04). 

The requirements are strongest for bright high-amplitude coherent oscillators. Considering a $m_{V}=4$ star with an amplitude of $10\%$ relative variability and a period of a few hours, target values have been set to 5 msec over the course of an observing sector for points (1) and (2), while the target value is 0.5 sec for point (3). For a solar-like oscillator the times should be accurate over a period of 10 days to better than 1 sec (3 sec for a red giant oscillator).

In this analysis we consider points (2) and (3) of the above, and refer to SAC\_TESS\_0002\_5 for more details \citep[see also][]{Montgomery1999}.

\section{Observations and data analysis}
\label{sec:Data}

\subsection{Ground-based photometry}

The ground-based observations presented in this work were collected using mainly the 2.15 meter telescope, {\it Jorge Sahade} (henceforth, CASLEO-2.15, programs JS-2018B-14, JS-2019A-02) and to a lesser extent the 0.6 meter telescope {\it Helen Sawyer Hogg} (henceforth, CASLEO-0.60, Director's Discretionary Time). Both telescopes are located at the Argentinian Complejo Astron\'omico El Leoncito (CASLEO). For CASLEO-2.15 we used a Roper Scientific model VersArray 2048B camera with a charge coupled device (CCD) detector (manufactured by Princeton Instruments) to collect the photometry. The imaging area is 2048$\times$2048 pixels, where each pixel is 13.5$\times$13.5 $\mu$m. The CCD is sensitive to wavelengths between 300 and \mbox{1000 nm}. To reduce dark current, the camera is cooled with liquid nitrogen and kept at approximately $-120$ degrees Celsius. With the mounted focal reducer, the circular, unvignietted field-of-view has a diameter of ${\sim}$9 arcminutes.  CASLEO-0.60 has a SBIG STL-1001E CCD, which is exclusively used for photometry. The imaging area is 1024$\times$1024 pixels, with a pixel size of 24$\times$24 $\mu$m. The CCD is sensitive to wavelengths between 400 and \mbox{1000 nm}, and is cooled down with a Peltier system. The telescope doesn't suffer vignetting, so the total fiel of  view is 9.26$\times$9.26 arcminutes. All our observations were performed using an \textit{R} filter, with an effective central wavelength, $\lambda_o$, of 635 nm and a full-width at half maximum (FWHM) of 107 nm. The main reason for this choice was to use a filter with a transmission response as similar as possible to the transmission response of TESS ($\lambda_o$ = 785 nm, FWHM = 400 nm), minimizing differences in the light curves associated with the wavelength-dependent stellar limb darkening. For a better overall photometric quality, this filter also circumvents the large telluric contamination around the \textit{I}-band. Contrary to TESS's constant $120$-sec cadence, the exposure time of the ground-based light curves depends mainly on the brightness of the star of interest, the altitude of the star during observations, and the photometric quality of the night during observations. In consequence, during an observing run we adjusted the exposure time so that the peak of the target point spread function was kept at around half the dynamic range of the CCD. This choice allows for an adequate compromise between linearity and good signal.

To achieve high precision photometry from the ground, we observed with the telescopes slightly defocused \citep{Kjeldsen1992,Southworth2009}. The achieved photometric precision per observing run is listed in column 4 of Table~\ref{tab:ObsCond}, along with other quantities derived from our observations.

\begin{table*}[ht!]
  \caption{\label{tab:ObsCond} Parameters derived from our ground-based observations. From left to right: the TESS Input Catalogue (TIC) and name of the observed target; the magnitude of the target in the TESS bandpass, m$_\mathrm{TESS}$; the date corresponding to the beginning of the local night; the telescope performing the observations; the standard deviation of the residual light curves in parts-per-thousand (ppt), $\sigma_{\rm res}$; the number of data points per light curve, $N$; the average cadence in seconds, CAD; the total observing time, $\Delta T_{\rm tot}$, in hours; the airmass range, $\chi_{\rm min,max}$, showing minimum and maximum values, respectively; the eclipse coverage, EC; a parameter to account for correlated noise, $\beta$ (see Section~\ref{sec:corrnoise}); and the derived time shifts, $\Delta t$, in seconds as compared to TESS data. The letter code specifying the eclipse coverage during each observation is as follows: O: out of eclipse, before ingress. I: ingress. B: bottom. E: egress. O: out of eclipse, after egress. Following the name the letter code (C) corresponds to the eclipses that have ground observations contemporaneous with TESS; (CG) corresponds to the eclipses that have contemporaneous observations between two ground-based stations; (NC) correspond to those that do not have contemporaneous observations, but a primary or secondary eclipse is clearly observed.}
    \centering
    \scalebox{0.92}{\hspace{-2cm}
    \begin{tabular}{l c c c c c c c c c c c}
    \hline \hline
    TIC/Name               & m$_\mathrm{TESS}$   &   Date           &  Telescope   &  $\sigma_{\rm res}$  &$N$&  CAD  & $\Delta T_{\rm tot}$ & $\chi_{\rm min,max}$ & EC & $\beta$ & $\Delta t$ \\
                           &                        &   yyyy.mm.dd      &              &  (ppt)               &   & (sec) & (hours)              &                      &    &         &   (seconds)   \\
    \hline
    69819180/KX Aqr (CG)   & 7.66  &  2018.06.07      & CASLEO-2.15  &  4.2  &   297  &  51.8  &  4.27  &  1.01,1.96  &  -IBE-  &  1.05  & 17 $\pm$ 138 \\
    69819180/KX Aqr (CG)   & 7.66  & 2018.06.07      & CASLEO-0.60  &  8.3  &   460  &  35.6  &  4.55  &  1.01,1.69  &  -IBE-  &  1.05  & 17 $\pm$ 138 \\
    349797905/NV Tel (NC)  & 9.66  & 2018.07.08      & CASLEO-2.15  &  1.3  &   343  &  34.9  &  3.32  &  1.14,1.53  &  -IBE-  &  1.03  & 302 $\pm$ 276 \\
    54018297/WY Cet (C)    & 8.85 & 2018.09.27      & CASLEO-2.15  &  7.9  &  1719  &  14.6  &  6.98  &  1.09,1.87  &  OI---  &  1.98  & 52 $\pm$ 77 \\
    9945183/VV Eri (C)     & 11.29 & 2018.10.30      & CASLEO-2.15  & 10.1  &   430  &  41.6  &  4.97  &  1.07,1.35  &  -IBEO  &  1.09  & -2.5 $\pm$ 10 \\
    220402294/BD Dor (C)   & 11.22 & 2018.11.10      & CASLEO-2.15  &  2.5  &   851  &  28.9  &  6.84  &  1.10,2.10  &  OIBEO  &  1.98  & -29 $\pm$ 26 \\
    260161144/AO Pic (C)   & 9.11   & 2018.12.10      & CASLEO-2.15  &  6.2  &  1462  &  18.4  &  7.45  &  1.11,1.58  &  OI---  &  1.46  &  7 $\pm$ 12 \\
    220402294/BD Dor (NC)  & 11.22 & 2018.12.11      & CASLEO-2.15  &  2.8  &   955  &  28.4  &  7.53  &  1.10,1.50  &  OIBEO  &  2.54  & 37 $\pm$ 15 \\
    80659292/AW Vel (C)    & 10.32 & 2019.01.25      & CASLEO-2.15  &  6.7  &  1280  &  21.7  &  7.73  &  1.02,1.55  &  OIBEO  &  1.86  &  30 $\pm$ 4 \\
    260161144/AO Pic (C)   & 9.11   & 2019.01.26      & CASLEO-2.15  &  5.4  &   351  &  58.7  &  5.72  &  1.35,1.89  &  OIBE-  &  1.39  & -9 $\pm$ 11 \\
    80659292/AW Vel (C)    & 10.32 & 2019.01.27      & CASLEO-2.15  &  1.7  &  1109  &  24.1  &  7.41  &  1.08,1.45  &  OIBEO  &  2.88  & -1 $\pm$ 8 \\
    220402294/BD Dor (C)   & 11.22 & 2019.03.24      & CASLEO-2.15  &  2.2  &   445  &  22.8  &  2.75  &  1.15,1.25  &  -IBE-  &  1.01  &  8 $\pm$ 14 \\
    219373406/X Pic (C)    & 10.51 & 2019.04.19      & CASLEO-2.15  &  5.3  &   288  &  21.7  &  1.73  &  1.33,1.85  &  -IBE-  &  1.03  & 2 $\pm$ 8 \\
    331183881/V636 Cen (NC)& 8.06  & 2019.05.20      & CASLEO-0.60  &  3.9  &   313  &  66.6  &  5.79  &  1.05,1.43  &  OIBEO  &  1.01  & -34 $\pm$ 15 \\
    41561453/RR Nor (C)    & 10.19 & 2019.05.23      & CASLEO-0.60  &  3.8  &   349  &  52.1  &  5.05  &  1.09,1.23  &  -IBE-  &  1.83  & -32 $\pm$ 7 \\
    214716930/TV Nor (C)   & 8.78  & 2019.06.07      & CASLEO-2.15  &  3.1  &  2188  &  12.4  &  7.58  &  1.06,1.53  &  OIBE-  &  1.01  & -30 $\pm$ 10 \\
    41561453/RR Nor (C)    & 10.19 & 2019.06.09      & CASLEO-2.15  & 11.6  &   811  &  17.0  &  3.83  &  1.12,1.70  &  OI---  &  1.23  &  86 $\pm$ 60 \\
    349797905/NV Tel (NC)  & 9.66  & 2019.07.17      & CASLEO-2.15  &  5.9  &   484  &  34.8  &  4.69  &  1.04,1.77  &  -IBE-  &  3.12  & -224 $\pm$ 216 \\
    \hline
  \end{tabular}}
\end{table*}

The ground-based data are reduced and the light curves are constructed by means of the {\it Differential Photometry Pipelines for Optimum Lightcurves}, DIP$^2$OL. A full description of DIP$^2$OL can be found in \citet{vonEssen2018}. In brief, the first component of the pipeline is based on \texttt{IRAF}'s command language \citep{iraf1993}, and it does aperture photometry. First, normal calibration sequences take place, depending on the availability of bias, darks and flatfield frames. The reduction continues with cosmic ray rejection and posterior alignment of the science frames. Afterwards, reference stars within the field are automatically chosen, usually of similar brightness to the target star to minimize the noise in the differential light curves \citep{Howell2006}. Photometric fluxes and errors are measured for all stars with different apertures, usually dividing the range from 0.5 to 3 times the nightly averaged FWHM in ten, and for each of these we use three different background rings. In this work we do not detrend the data, as the eclipses are deep (usually $\Delta$Flux$\sim$50-80\%). Instead, we treat their noise as explained in Section~\ref{sec:corrnoise}. The second part of DIP$^2$OL is written in \texttt{Python}. The routine produces several light curves using different combinations of reference stars. The final differential light curve is the unweighted sum of the flux of the target star divided by the sum of the unweighted fluxes of the reference stars that produced the light curve with the smallest point-to-point scatter. The pipeline repeats this process per aperture and sky ring. The code outputs the time in Julian dates shifted to the center of each exposure, the differential fluxes, photometric error bars and the detrending quantities that are ignored in this work. In particular, a high degree and unphysical time-dependent polynomial is fitted to the light curve through least-squares minimization. A residual light curve is constructed producing the difference between the final light curve and the best-fit polynomial. From this residual light curve we compute the standard deviation, and we use this value to enlarge the photometric error bars, so that their average is at the same level of this standard deviation. With the light curves fully constructed, we convert the time stamps from Julian dates to Barycentric Julian dates, BJD$_\mathrm{TDB}$, using the web tool provided by \cite{Eastman2010}. The stars listed in Table~\ref{tab:ObsCond} are used in three different ways. Those having a (C) correspond to the eclipses that have contemporaneous observations with TESS, while those with a (NC) do not have contemporaneous observations, but an eclipse is clearly observed. Both were used to compute potential time offsets (Section~\ref{sec:time_offset}). Those with an (CG) correspond to eclipses that have contemporaneous observations from the ground, and were used to test the timings between CASLEO-2.15 and CASLEO-0.60 (Section~\ref{sec:time_test}). 

\subsection{Correlated noise for the ground-based light curves}
\label{sec:corrnoise}

Several problems arise when observing stars from Earth as compared to space. Fluctuations in the atmosphere causing an abrupt dimming of the star, clouds suddenly appearing and poor tracking of the observed stellar fields are only some of the many nuisances that have to be overcome in order to obtain accurate, reliable data. Thus, different techniques have been developed in order to eliminate these nuisances. As the eclipses are deep and the target stars are relatively bright, we do not detrend the photometry, but rather increase the individual photometric uncertainties by the so-called $\beta$ factor to account for correlated noise \citep{Pont2006,Carter2009}. 

In order to compute the $\beta$ factor, as described in \cite{vonEssen2013}, we first compute residuals by fitting a high order, non-physical, polynomial to the ground-based light curves. Then, we divide the residuals into $M$ bins of $N$ averaged data points. This average accounts for changes in exposure time that might be needed to compensate for changes in airmass or transparency during the observing runs. Due to the usual length of our ground-based data sets, we consider bins of four different lengths, namely 10, 15, 20, and 25 minutes.

In general, if the data have no correlated noise then the noise in the residuals should follow the expectation of independent random numbers:

\begin{equation}
  \hat{\sigma}_N = \sigma N^{-1/2}[M/(M-1)]^{1/2}\ ,
\end{equation}

\noindent where $\sigma$ is the standard deviation of the unbinned residual light curve, and $\sigma_N$ corresponds to the standard deviation of the data binned with $N$ averaged data points per bin:

\begin{equation}
  \sigma_N = \sqrt{\frac{1}{M}\sum_{i = 1}^{M}\left(\langle\hat{\mu}_i\rangle - \hat{\mu}_i\right)^2}\ .
\end{equation}

\noindent In the equation above, $\hat{\mu_i}$ corresponds to the mean value of the residuals per bin ($i$) and $\langle\hat{\mu_i}\rangle$ is the mean value of the means. $\beta$ is computed averaging $\beta_N$ = $\hat{\sigma}_N$/$\sigma_N$, computed in the time bins mentioned before. When we found $\beta$ to be larger than 1, we enlarged the individual photometric errors of the ground-based light curves by this factor, and only then we carried out the determination of the individual mid-eclipse times.

\subsection{TESS data}

During the first thirteen sectors, the eclipsing binary systems comprising our timing verification list were observed with a cadence of 120 seconds. For the 120-sec cadence data we adopted the PDCSAP light curves provided by the Science Processing Operations Center \citep[SPOC;][]{Jenkins2016} pipeline in the Target Pixel Files (TPFs)\footnote{https://archive.stsci.edu/missions/tess/doc/EXP-TESS-ARC-ICD-TM-0014.pdf}, which were downloaded from the TASOC database\footnote{tasoc.dk}. Only for \mbox{BD Dor} was it necessary for us to create custom light curves for Sectors 2-5, as during these sectors \mbox{BD Dor} was incorrectly associated with the target \mbox{TIC 220402290}. As this target lies only ${\sim}3$ pixels away from the correct target, \mbox{TIC 220402294}, both stars were included in the photometric aperture. The eclipses of \mbox{BD Dor} were in consequence observable, but were highly diluted by the contribution of \mbox{TIC 220402290} to the total flux. This missidentification, that was also found in other catalogues, was reported by our group to the Centre de Données astronomiques de Strasbourg and corrected. As previously mentioned, given the proximity of the two stars they are both contained in the TPFs for \mbox{TIC 220402290}, so we simply defined a new aperture around the correct target. In later Sectors, \mbox{BD Dor} is correctly associated with \mbox{TIC 220402294}.

\section{Determination of the mid-eclipse times}
\label{sec:Mid-Eclipses}

Depending on the specific binary system observed, and thus the spectral type of the stars, their relative sizes and their mutual distances, the overall shape of the eclipses will significantly change from one system to the other. One model to determine mid-eclipse timings would not accommodate a wide range of difference eclipse shapes. To overcome this we have developed three different ways to extract the eclipse timings of ground and space-based data. 
The first involves the use of a time-dependent second order polynomial, the second an inverted Gaussian function, and the third is similar to a cross-correlation between two contemporaneous light curves. The first two techniques are specified in Section~\ref{sec:functions}, while the cross-correlation method is detailed in Section~\ref{sec:crosscorr}. Regardless of the model used, timing offsets between TESS and ground-based data are computed in three ways. From the three results, we always report the one with the smallest difference.

\subsection{Model functions for the mid-eclipse times}
\label{sec:functions}

A method for computing accurately the mid-eclipse times was first given by \cite{Kwee1956}. Following their approach, our first model corresponds to a time-dependent, second order polynomial, 

\begin{equation}
 f(t) = a t^2 + b t + c,\,
\end{equation}

\noindent where $a$, $b$ and $c$ are the fitting parameters. Here, the mid-eclipse time is computed as \mbox{$T_o = -b/2a$}, and its associated error is computed from standard error propagation. 

The second model is an inverted Gaussian function,

\begin{equation}
 g(t) = \beta - \alpha e^{-\frac{(t - \mu)^2}{2\sigma^2}},\,
\end{equation}

\noindent where $\alpha$, $\beta$, $\mu$ and $\sigma$ are the fitting parameters, and the mid-eclipse time is computed as \mbox{$T_o = \mu$}.

\subsection{Optimum window around mid-eclipse to derive accurate timings}
\label{sec:crosscorr}

\begin{figure}[ht!]
    \centering
    \includegraphics[width=.5\textwidth]{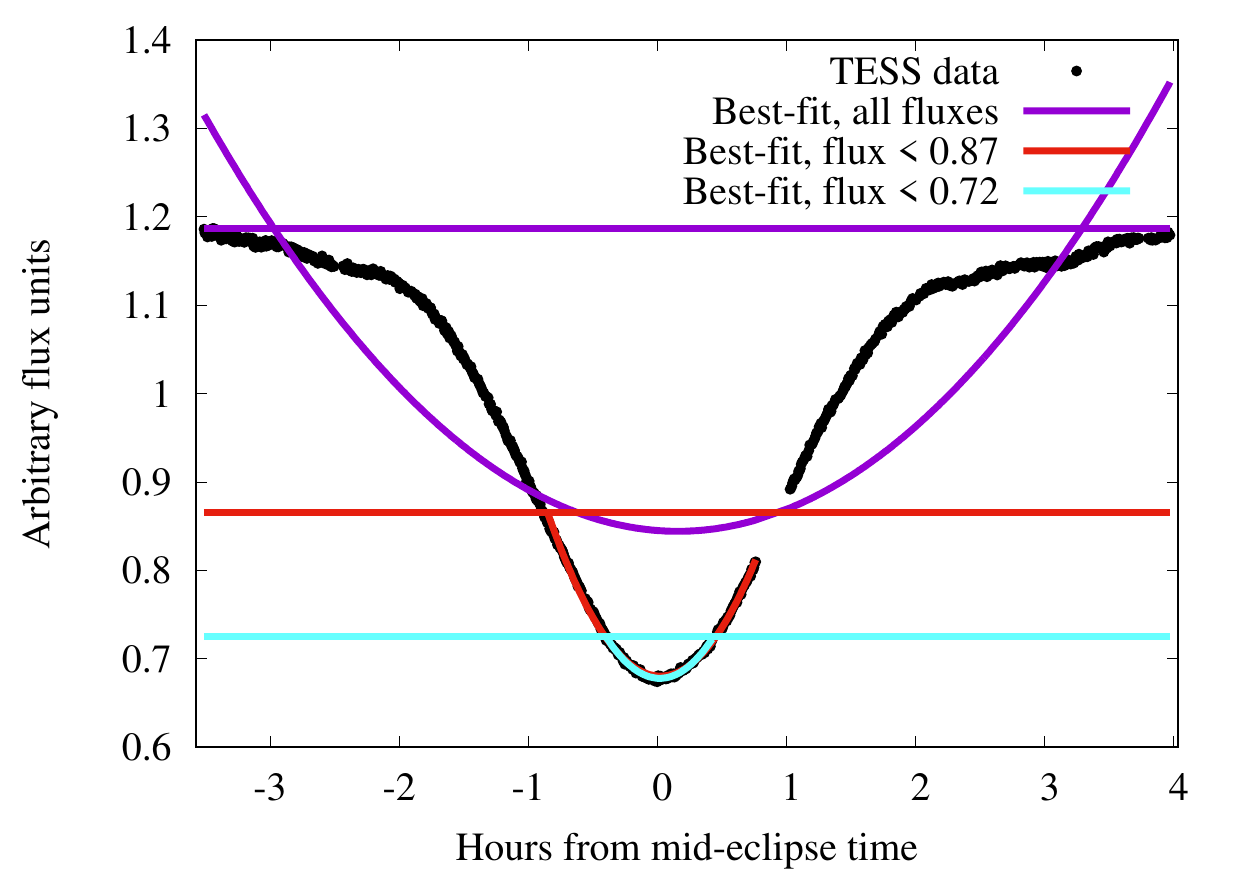}
    \caption{\label{fig:sorting} Flux in arbitrary units as a function of hours from the time of mid-eclipse. The black points show one eclipse of \mbox{BD Dor} from TESS observations. Overplotted are three second-order polynomial fits to all the visible flux in violet, flux values lower than 0.87 in red, and flux values lower than 0.72 in cyan. Dashed horizontal lines indicate these levels and are placed to guide the eye.}
\end{figure}

While TESS data are largely continuous within a sector, ground-based observations face other challenges, mainly imposed by the diurnal rotation of the Earth and the cloud coverage. In consequence, the coverage from CASLEO does not resemble that from TESS. In some cases, the eclipse coverage is asymmetric, in some eclipses the instant of minimum flux is missing, and in some others there are gaps without data. This inconvenient coverage will have an impact in the precision of the derived timings. To overcome this, before calculating the mid-eclipse times we sort the data to find an optimum number of data points (and thus, eclipse coverage) that best match our models. The sorting function will gradually remove data points with a flux larger than a specified value. After each round of trimming, the remaining data points are fitted with our models (Section~\ref{sec:functions}). The gradual chopping starts at the maximal observed flux, $f_\mathrm{max}$, and ends at \mbox{$f_\mathrm{min}$ + $0.1 (f_\mathrm{max} - f_\mathrm{min})$}, where $f_\mathrm{min}$ corresponds to the minimum observed flux value, and 0.1 is user specified. The reason why the sorting does not reach $f_\mathrm{min}$ is to ensure that the amount of fitting parameters does not exceed the number of data points. An example of the use of this sorting strategy can be seen in Figure~\ref{fig:sorting}, where three different fits are shown for three different chopping values. As we are only determining the optimum window that best match our models, the fits are carried out by means of a simple least-squares minimization. After performing each fit, we compute the reduced chi-squared, $\chi^2_\mathrm{red}$, considering at each step the changing number of data points. The final eclipse coverage of the ground-based light curves used to determine mid-eclipse times is the one corresponding to a $\chi^2_\mathrm{red}$ value equal to (or close to) one. In the figure, the eclipse coverage that best matches our model lies between the blue and cyan lines.

\subsection{Cross-correlation}
\label{sec:shift_squeeze}

Our third method resembles a cross-correlation between TESS and CASLEO data. Here, the proper time lag between data sets is determined by minimizing the sum of the squared residuals. This sum should approach zero as the correlation between the two data sets become larger. It is generally straightforward to visualize the method in the following way. As one data set is kept fixed through the entire process, the other set is shifted along the abscissa with a given time lag and a scaling is applied to the ordinate values to scale one light curve to the other. Due to the continuity of the TESS data the full eclipse is typically covered, unlike the ground-based light curves (see column 9 of Table~\ref{tab:ObsCond} for the eclipse coverage). We therefore always considered the TESS data as the data being shifted, because here one can better scale the light curve. The scaling is applied because the light curves may appear different due to limb darkening or due to errors while constructing them, such as aperture losses or intrapixel variations. The first step in the program is to center the eclipses around zero in the ordinate so that the applied scale will squeeze or stretch the light curve from TESS, and not simply multiply the flux by a given factor. This is done by calculating and subtracting the mean of each data set separately. With the data sets varying in size, the mean value is computed considering data points where both TESS and CASLEO observations are defined. A time lag is then applied to the TESS data and the flux is linearly interpolated and evaluated at the times of the ground-based data. The mean is then recalculated and subtracted once more from the TESS data, which is necessary to account for the potential change after interpolating to CASLEO's timings. We then proceed in calculating the sum of the squared residuals.
Both time lags and scaling factors are obtained from grids with sensible ranges: $\pm$60 seconds with a step of 1 second for the time lag, and $\pm$10\% variability with a step of 0.5\% for the scaling. For each combination of parameters the sum of squared residuals is computed. The final time lag is the one that minimizes the sum of squared residuals.

\subsection{Errors on the mid-eclipse times}
\label{sec:errors_Tos}

To compute reliable uncertainties for the mid-eclipse times determined from TESS and CASLEO data using the three approaches described in Section~\ref{sec:Mid-Eclipses}, we determine the timing uncertainties by fitting the data and models using a Markov-Chain Monte Carlo (MCMC) approach, as implemented in \texttt{PyAstronomy}\footnote{\label{note1}\url{https://github.com/sczesla/PyAstronomy}}, a collection of \texttt{Python} routines implemented in the \texttt{PyMC} \citep{Patil2010} and \texttt{SciPy} \citep{Jones2001} packages. The best-fit mid-eclipse times and their uncertainties are derived from the mean and standard deviation \mbox{(1-$\sigma$)} of the posterior distributions of the fitted parameters, which are drawn from $10^5$ iterations after carrying out a conservative burn-in of 20\% of the initial samples. This burn-in was determined from prior visual inspection of the chains. 

\section{Results}
\label{sec:results}

\subsection{Testing timings between CASLEO-2.15 and CASLEO-0.60}
\label{sec:time_test}

\begin{figure}
    \centering
    \includegraphics[width=.5\textwidth]{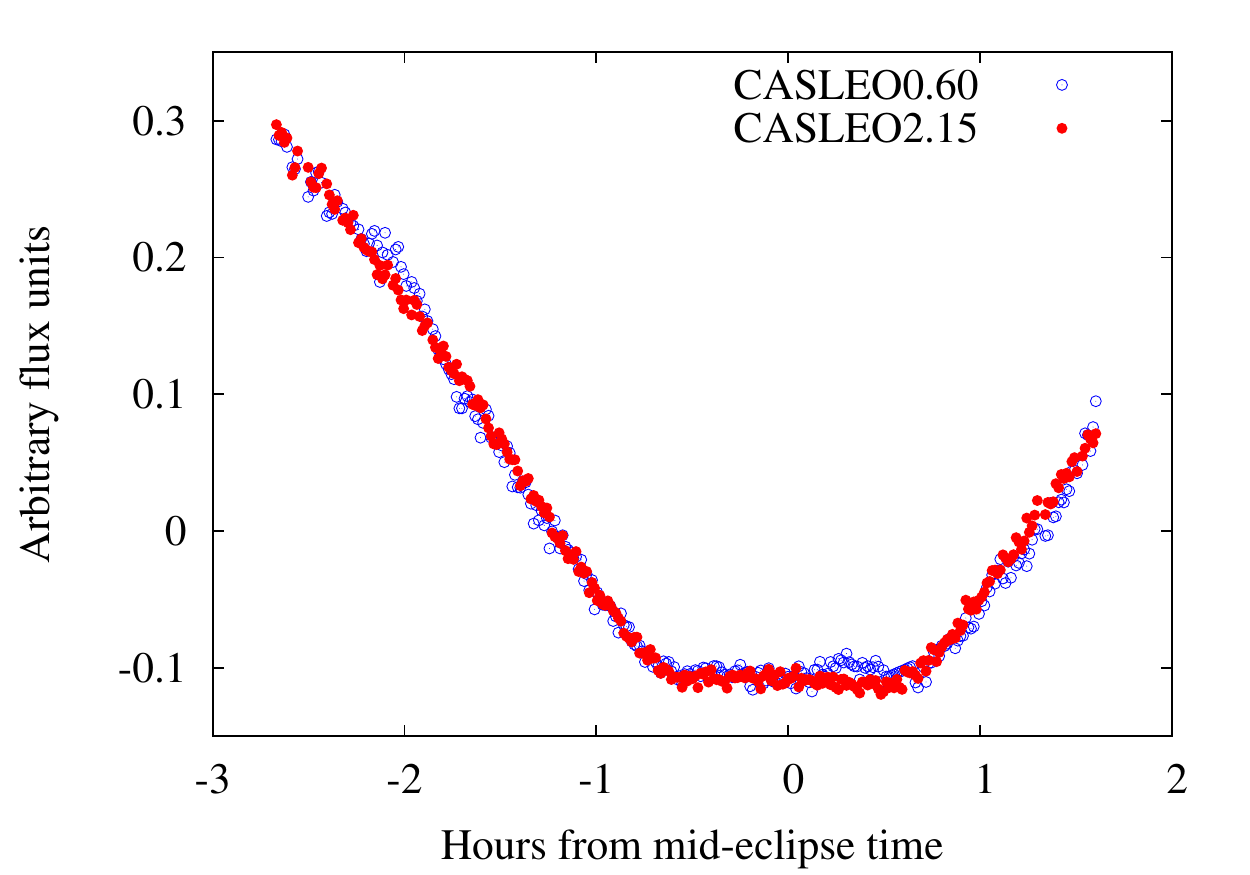}
    \caption{\label{fig:KX_Aqr} Ground-based eclipse observation of \mbox{KX Aqr} collected with CASLEO-2.15 (red filled circles) and with CASLEO-0.60 (blue empty circles). The figure displays arbitrary flux units as a function of hours from mid-eclipse time.}
\end{figure}

Verifying TESS timestamps from ground-based observations means that the TESS timestamps will be limited by the accuracy of the ground-based observatios. In consequence, the success of this technique relies on how accurately CASLEO's telescopes can report their own timestamps. Both telescopes collect the Universal time from two identical global positioning systems (GPS). The sidereal time is based on a micro-controller synchronized with the GPS that sends its timing to the Programmable Logical Controllers, which in turn are in charge of collecting the data. Despite the professional setup, we carried out an independent check of their timing resemblance. To do so, on the night of June 7, 2018, we observed the eclipsing binary \mbox{KX Aqr} contemporaneously with CASLEO-2.15 and CASLEO-0.60. The target was not observed by TESS. Figure~\ref{fig:KX_Aqr} shows CASLEO-2.15 data in red, and CASLEO-0.60 data in blue. The timing difference between the two data sets was obtained using the cross-correlation method described in Section~\ref{sec:shift_squeeze}. Its value, of \mbox{19 $\pm$ 85 seconds}, is consistent with zero at 1-$\sigma$ level. The large uncertainty, in this case, reflects the high noise in the CASLEO-0.60 light curve. A detailed description of the observations can be found in the first two lines of Table~\ref{tab:ObsCond}.

\subsection{Testing for a time offset}
\label{sec:time_offset}

\begin{figure*}[ht!]
    \centering
    \includegraphics[width=.33\textwidth]{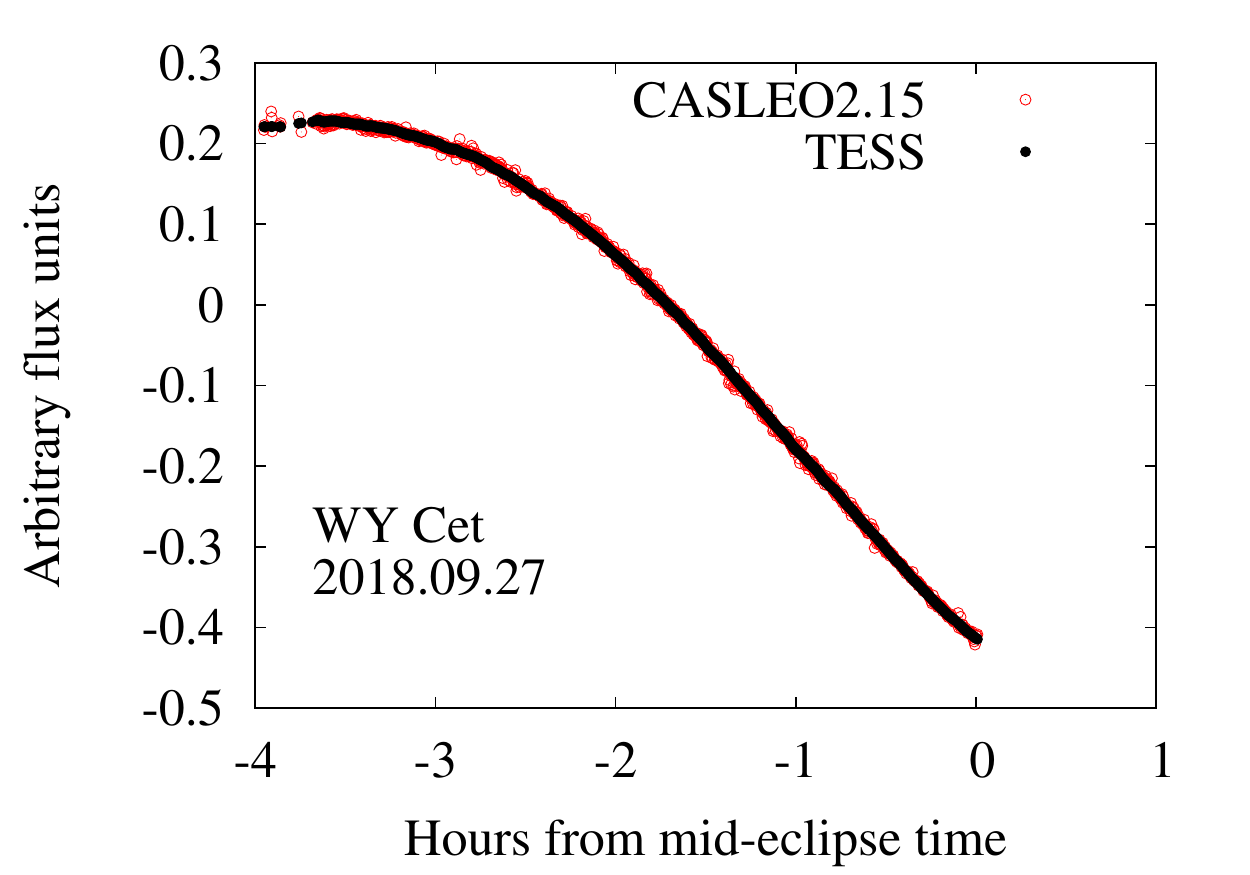}
    \includegraphics[width=.33\textwidth]{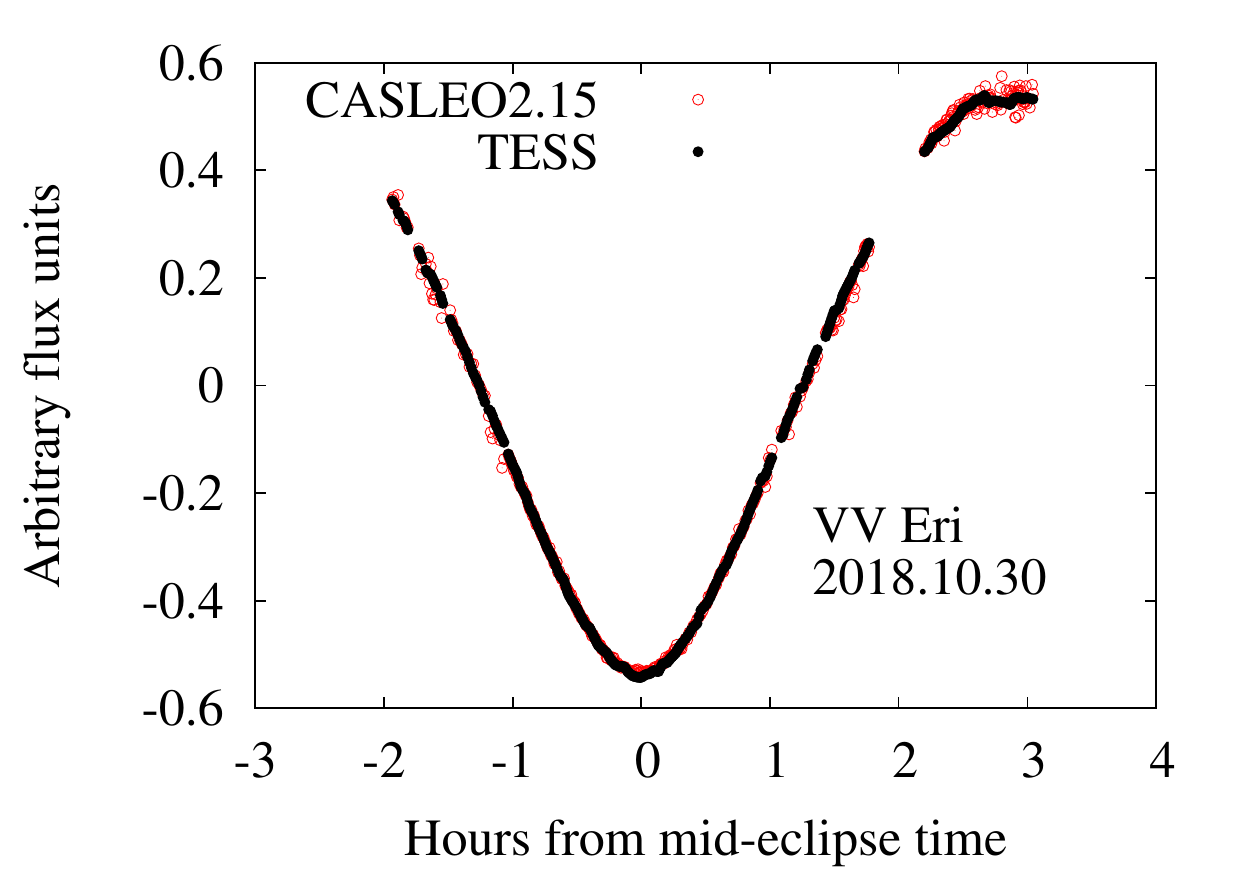}
    \includegraphics[width=.33\textwidth]{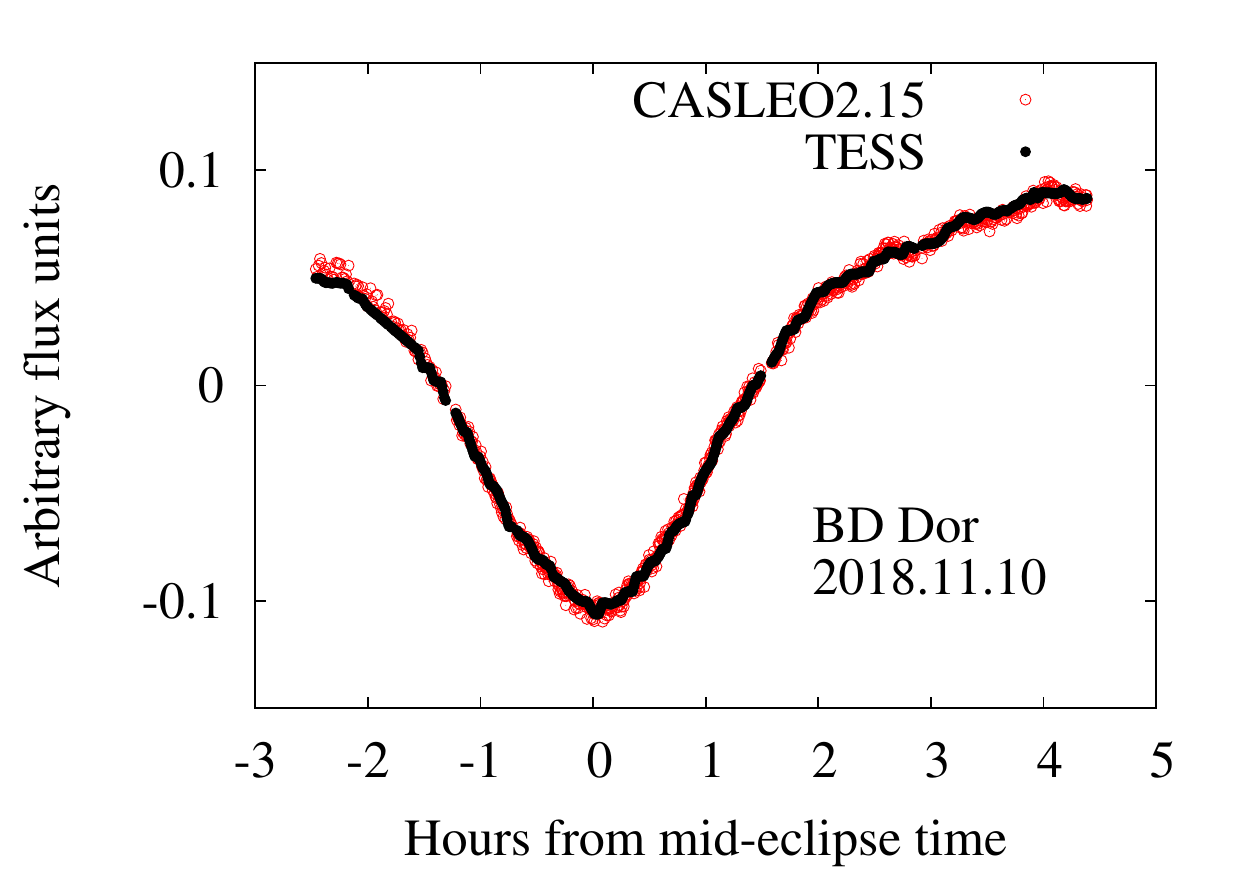}
    
    \includegraphics[width=.33\textwidth]{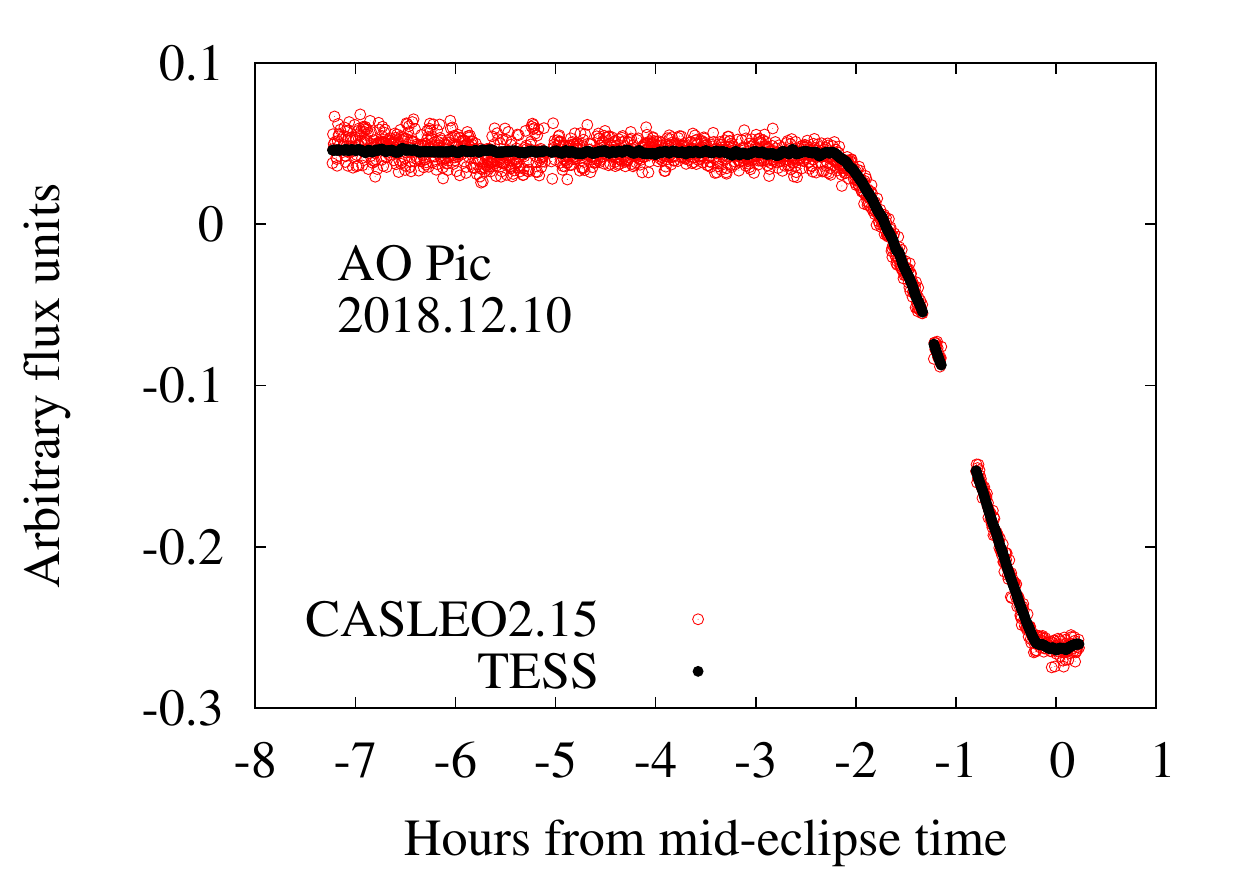}
    \includegraphics[width=.33\textwidth]{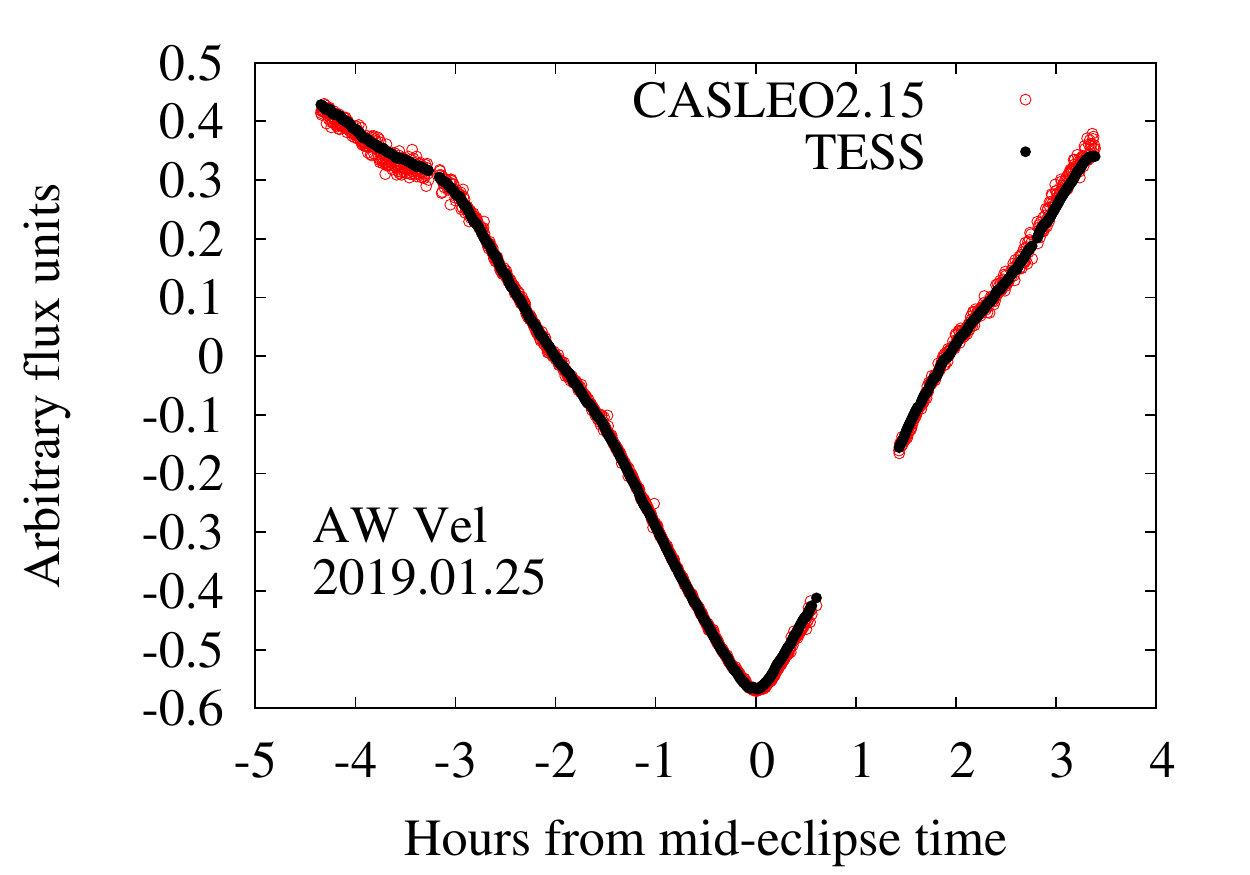}    
    \includegraphics[width=.33\textwidth]{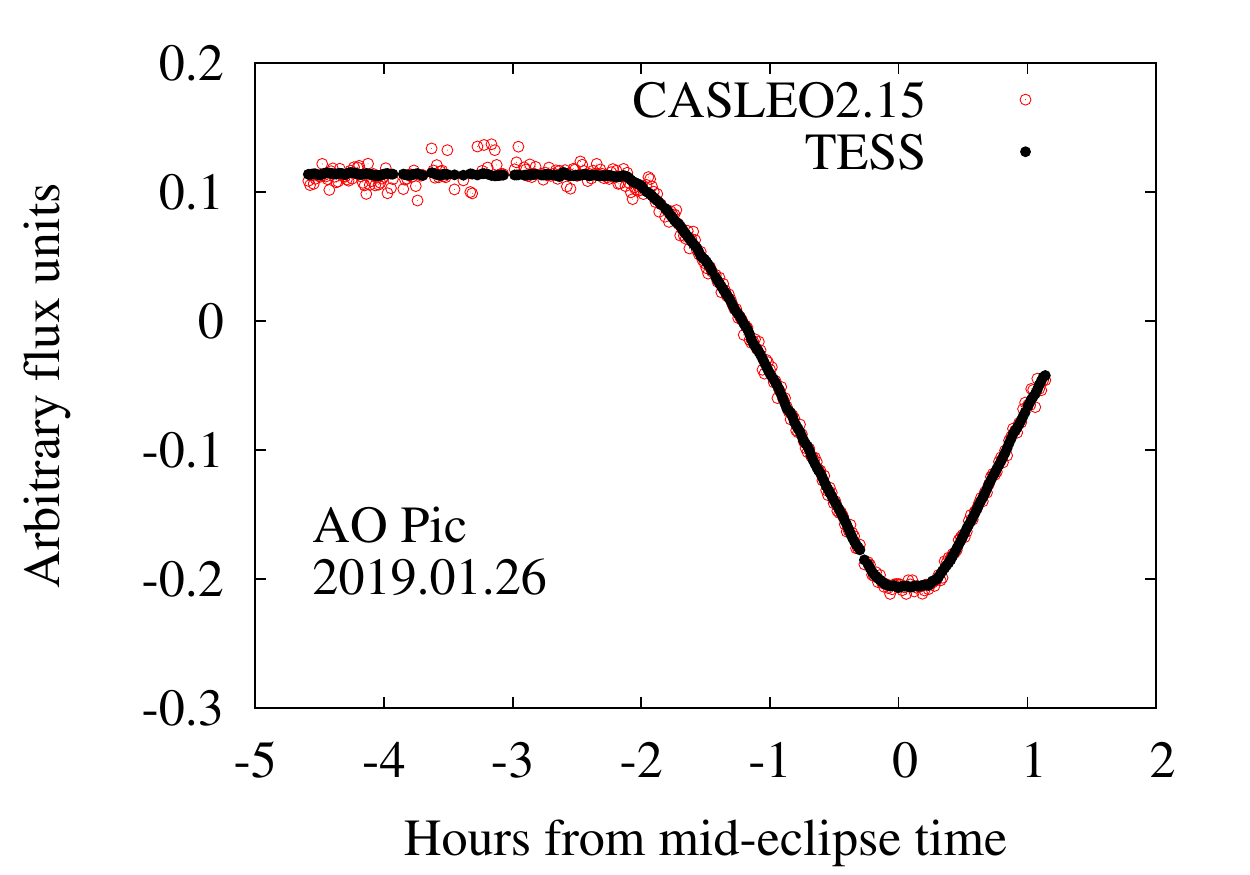}
    
    \includegraphics[width=.33\textwidth]{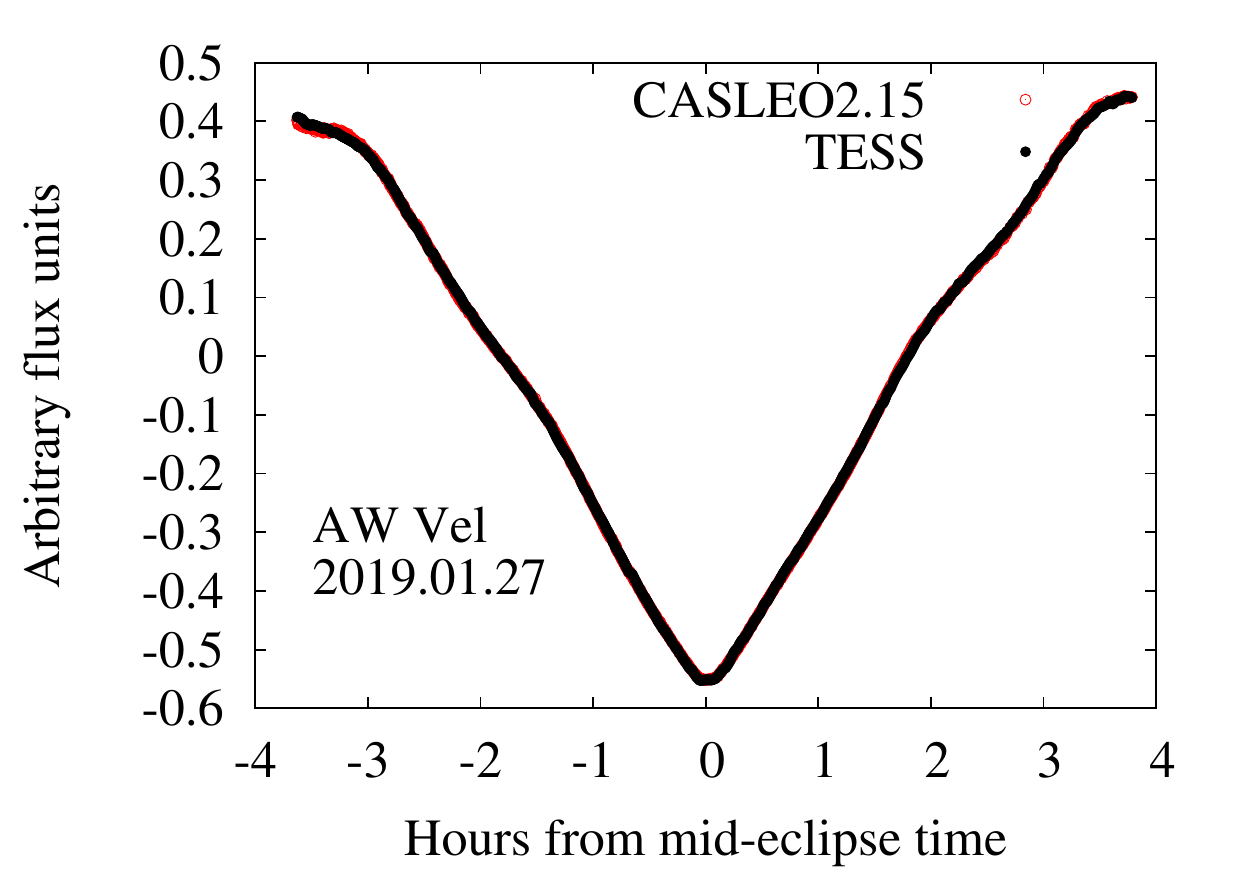}
    \includegraphics[width=.33\textwidth]{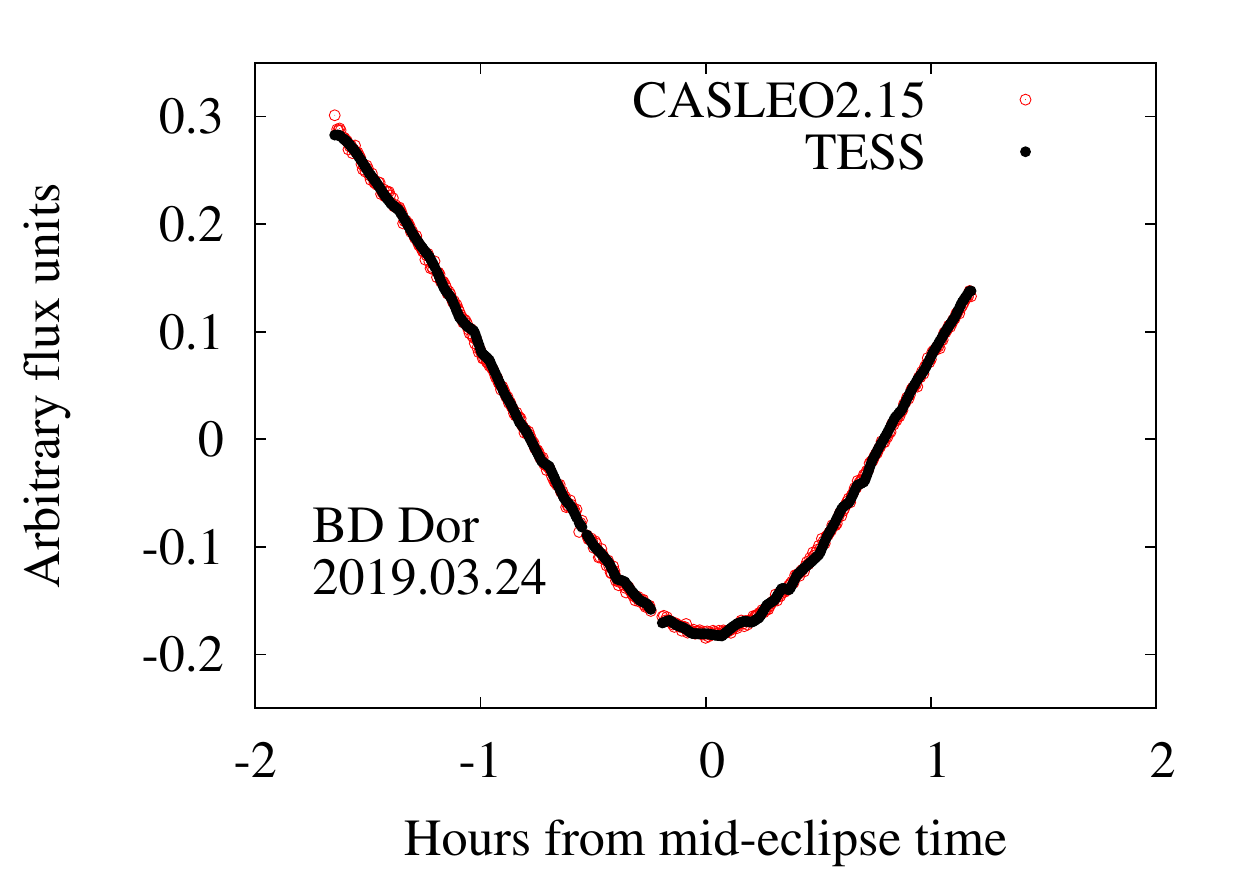}
    \includegraphics[width=.33\textwidth]{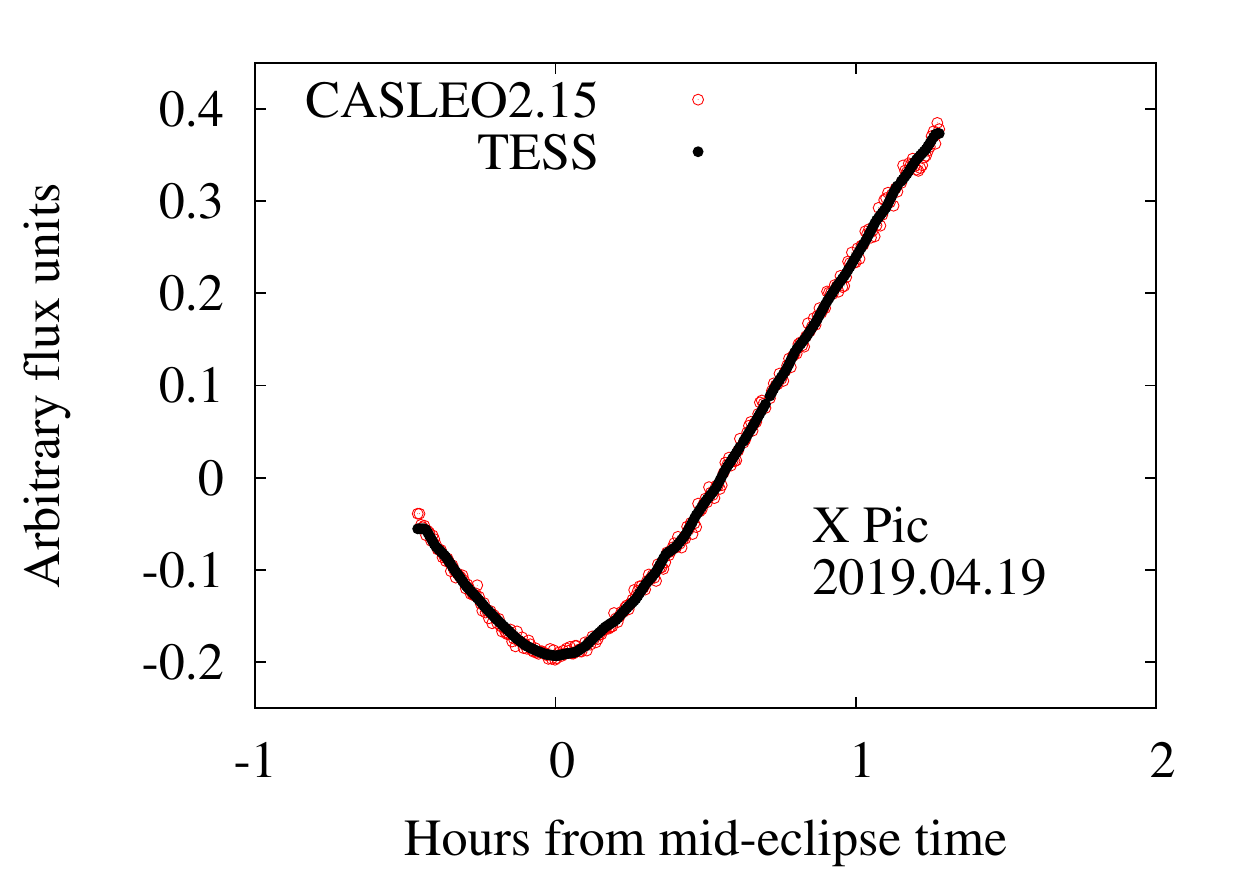}
    
    \includegraphics[width=.33\textwidth]{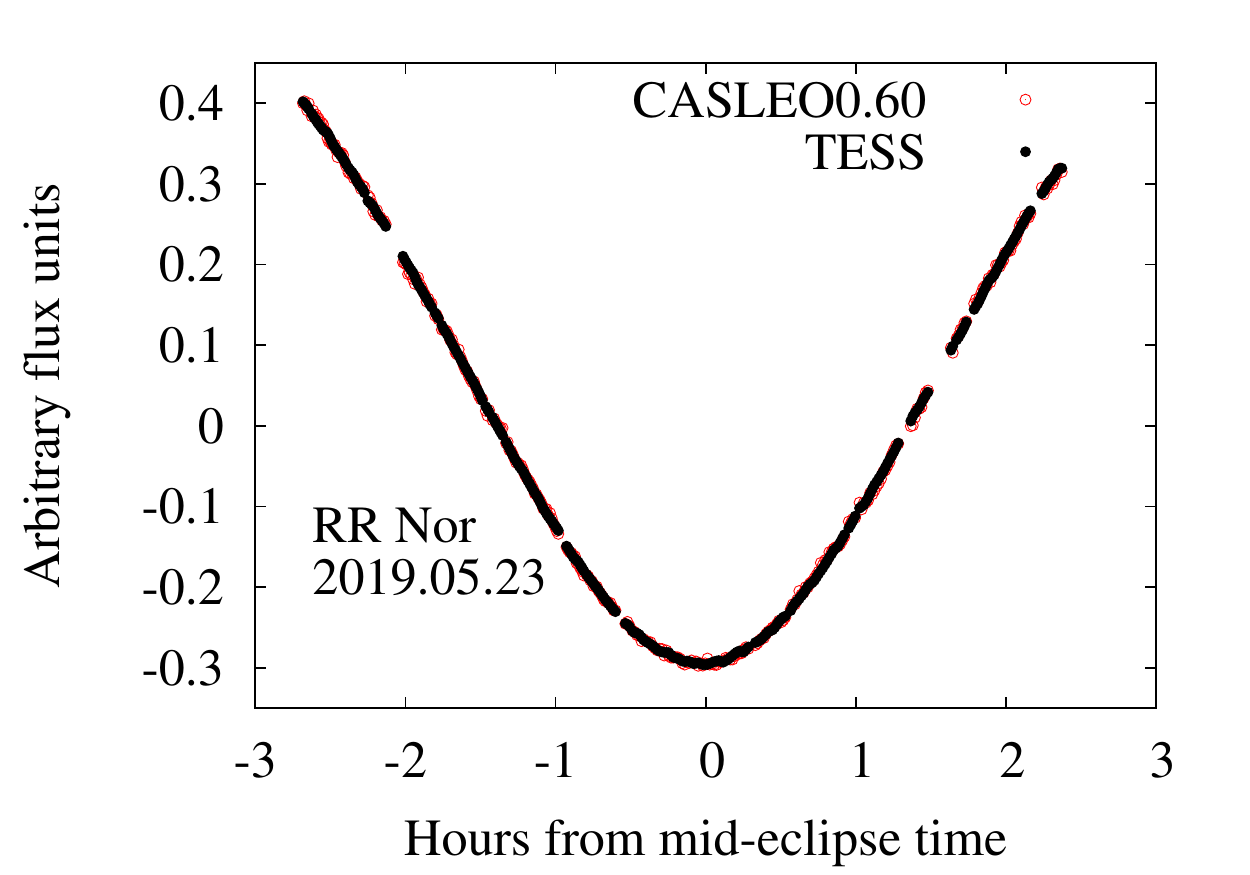}
    \includegraphics[width=.33\textwidth]{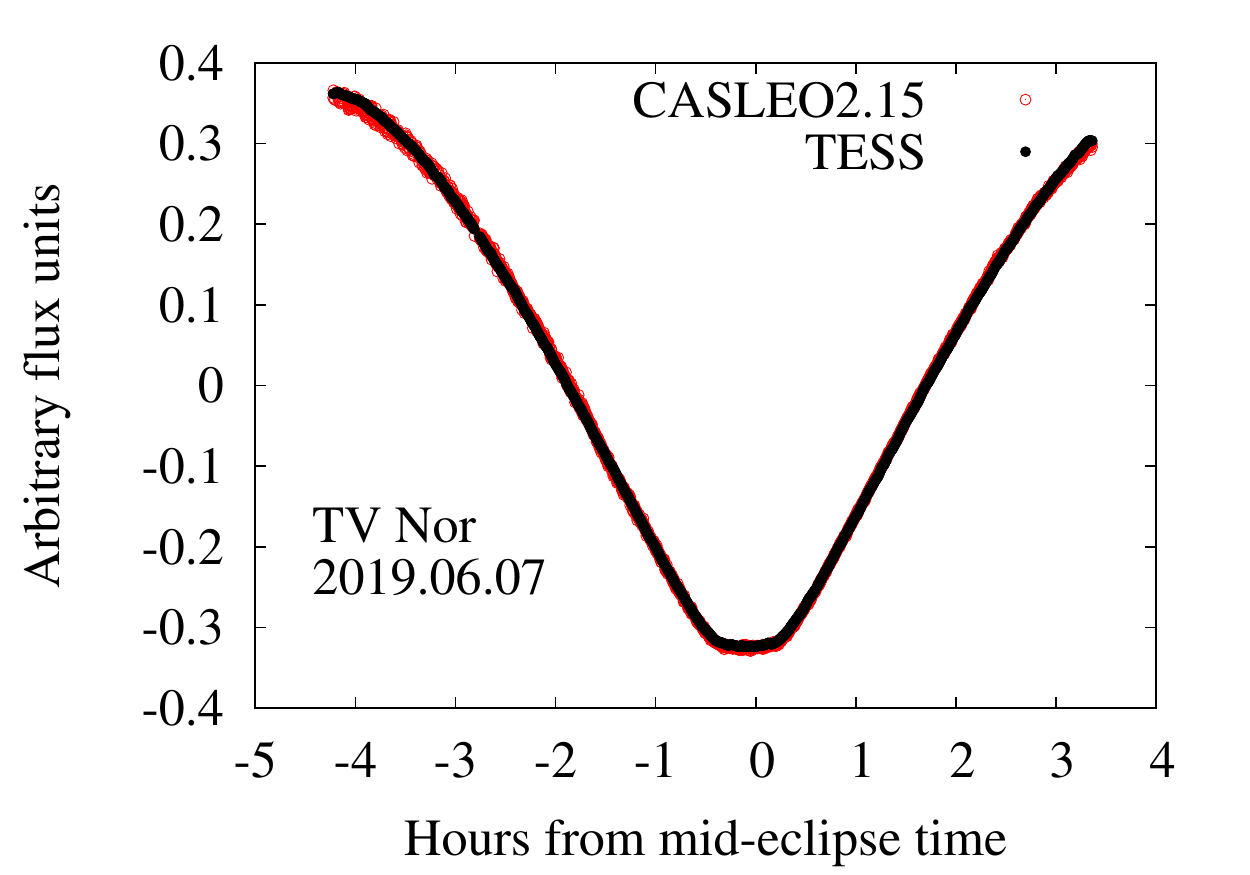}
    \includegraphics[width=.33\textwidth]{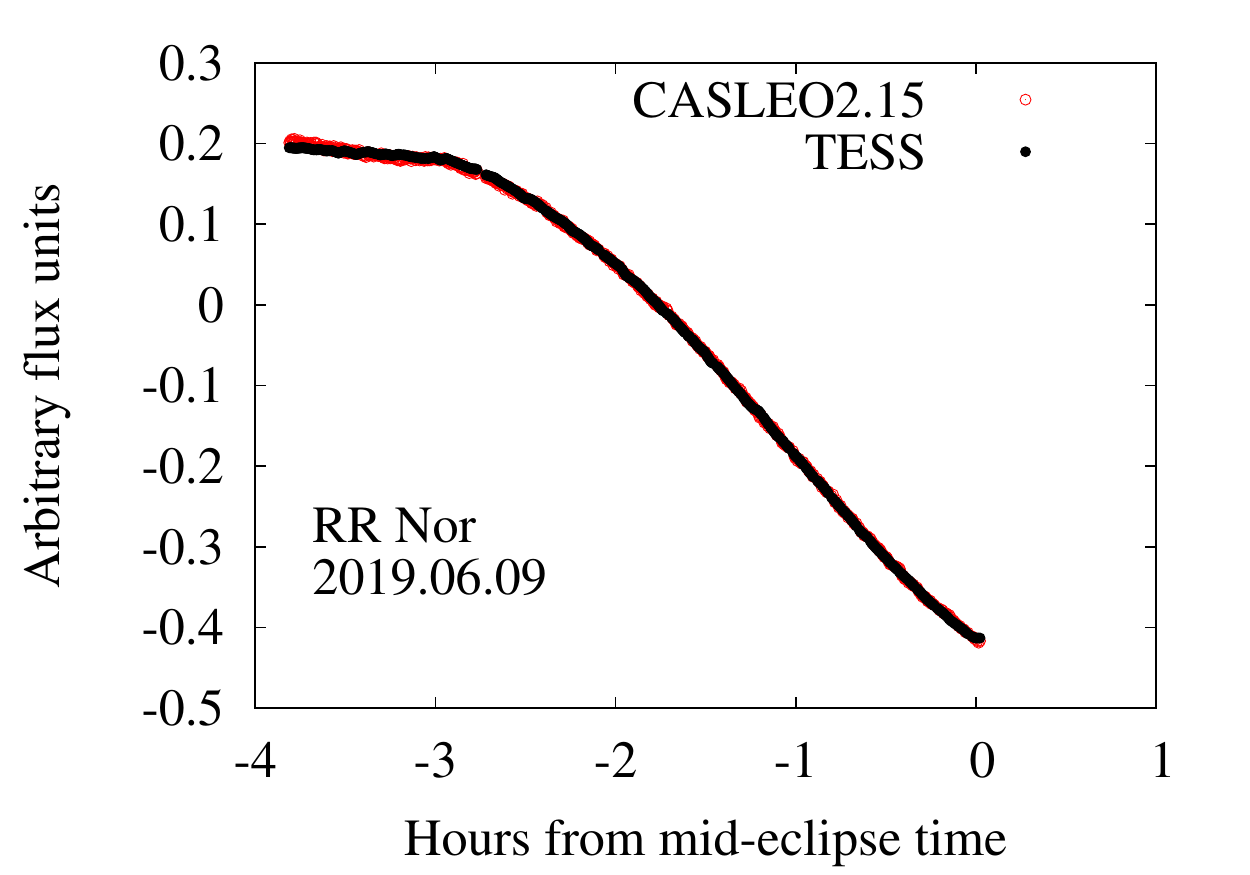}
    \caption{\label{fig:eclipses} Ground versus space-based photometry. The figures are arranged chronologically from left to right and top to bottom. All figures show arbitrary flux units as a function of the hours from mid-eclipse time. Overplotted are CASLEO-2.15 and CASLEO-0.60 data in red open circles, and TESS data in black filled circles. The data sets are shifted to maximize their overlap. The applied time shift is specified in the last column of Table~\ref{tab:ObsCond}. In each sub-figure we specify the name of the target, the telescope that performed the observations and the date corresponding to the beginning of CASLEO's observing night.}
\end{figure*}

Our work is based on the determination of mid-eclipse times of selected binary systems observed from TESS and from CASLEO's telescopes. Thus, it is expected that the times of minimum flux will occur simultaneously. This will not necessarily be observed if there is a time offset or a time drift in the clock on-board TESS. To determine a potential time offset of the TESS timestamps, we observed eclipses from several binary systems during TESS's first thirteen sectors. Several aspects reduced the number of good contemporaneous data sets. Some examples are outdated ephemerides, which produced inaccurate windows at which to observe from the ground, and poor weather conditions during observations leading to poor photometric quality in the derived light curves. As a consequence, not all the eclipses listed in Table~\ref{tab:ObsCond} have contemporaneous ground-space observations. Only the twelve specified with a (C) next to their names do. For each one of these eclipse observations, we computed the mid-eclipse times as detailed in Sections~\ref{sec:functions} and \ref{sec:shift_squeeze}, and the timing differences between TESS and ground (Table~\ref{tab:ObsCond}). Figure~\ref{fig:eclipses} shows the corresponding light curves. The errors in the Observed-minus-Calculated (O-C) points are computed from simple error propagation, taking into account the individual timing uncertainties. Averaging the timing differences computed from the twelve contemporaneous eclipses, the derived mean timing offset is \mbox{5.8 $\pm$ 2.5 seconds}. As some of our points in the O-C diagram have a larger offset and a corresponding large uncertainty, in order to properly take them into account our reported offset was obtained computing the weighted mean, and its uncertainty was derived from the standard error of the weighted mean. The timing differences are shown as black squares in Figure~\ref{fig:OC_TESS}. If no time offset exist the O-C points should be normally distributed with zero mean. Only recently, the TESS team discovered a time offset of 2 seconds\footnote{\url{https://archive.stsci.edu/missions/tess/doc/tess_drn/tess_sector_22_drn31_v01.pdf}}. By taking this offset into consideration, our results improve to \mbox{3.8 $\pm$ 2.5 seconds}, only 1.5-$\sigma$ away from zero. It is worth to mention that as of sector 20, the data products on the Mikulski Archive for Space Telescopes (MAST) are corrected by the 2 second offset.

If our derived mid-eclipse times are properly computed and don't show any systematic effect that arises purely from our procedures, they should follow a normal distribution. To assess this, we performed a Kolmogorov-Smirnov test \citep{KolmogorovS} in which we compared our TESS-ground timing differences against a normal distribution. The derived p-value of \mbox{p = 0.913} does not allow us to reject the null hypothesis that both distributions are the same. In addition, we used the best-fit orbital periods and mid-eclipse times of reference listed in Table~\ref{tab:ephemeris} to determine the timing differences between the observed mid-eclipse times corresponding to the four (NC) data sets and the corresponding ones computed from the ephemeris. Two of the O-C points are shown in Figure~\ref{fig:OC_TESS} in blue triangles, as the other two are off-range to allow for proper visual inspection. The derived timing offset is \mbox{2 $\pm$ 11 seconds}, consistent with zero at 1-$\sigma$ level.

\subsection{Testing for a time drift}
\label{sec:time_drift}

The time drift method relies solely on space-based data, exploiting the power of the continuous observations of all the short-period binary systems followed by TESS for the time verification. The advantage of this method is that it can be run without having ground-based data. The disadvantage is that care must be taken when trying to interpret the derived O-C diagrams. Even though a trend may occur, this does not necessarily stem from TESS timing-drifts. It could instead be a result of the physics in the system itself. Thereby, the same trend must occur in the O-C diagram for several binary systems. It will also not be possible to infer anything about the absolute timing with this method, as the TESS time is not compared to an outside source, so the only possible result from this method is an assessment of a potential drift in the times. 

To determine the potential time drift in TESS timings we proceed as follows. For each system we determined the individual mid-eclipse times by carrying out the mid-eclipse timing strategy presented in Section~\ref{sec:Mid-Eclipses}. From the individual mid-eclipse times we determined the orbital period and mid-eclipse time of reference per system. To compute the timing deviation compared to a constant period, we fitted the observed mid-eclipse times, T$_{o,i}$, to the expression:

\begin{equation}
  T_{o,i} = P \times E_i + T_0\,.
  \label{eq:nosigma}
\end{equation}

\begin{figure*}[ht!]
    \centering
    \includegraphics[width=.95\textwidth]{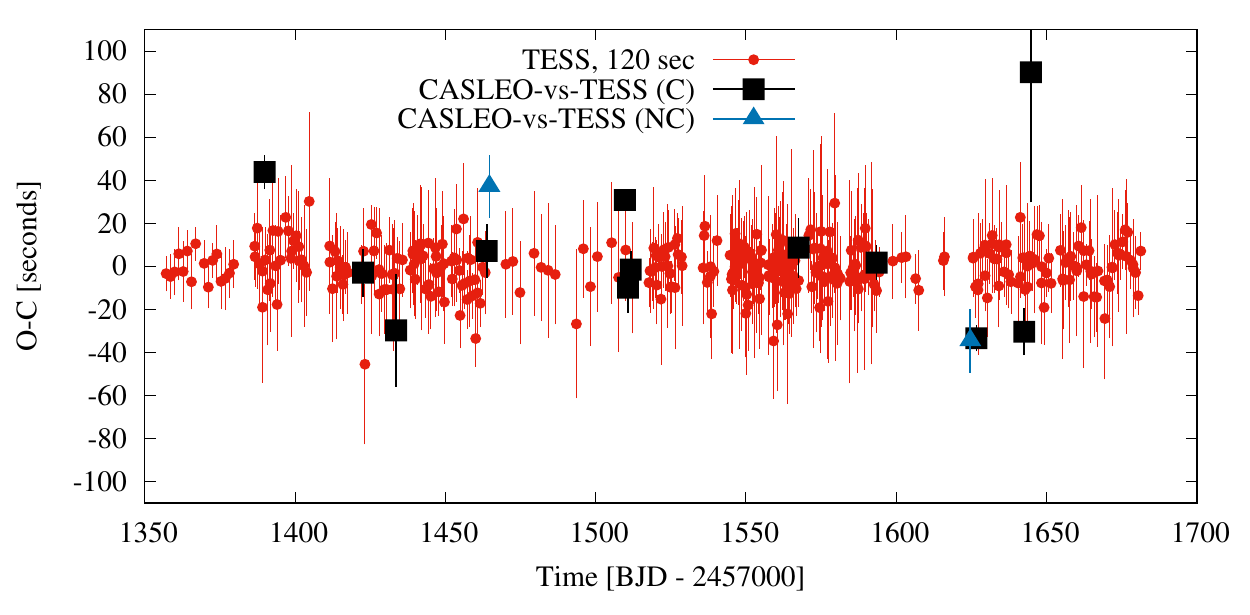}
    \caption{\label{fig:OC_TESS} O-C diagram of the 120-sec TESS data in red; timing differences between the contemporaneous TESS and CASLEO data in black squares, and the non-contemporaneous in blue triangles. Uncertainties are in all cases given at $1-\sigma$ level. The large uncertainty and offset of the last black square corresponds to RR Nor, and it is the product of a partial observation.}
\end{figure*}

\noindent Here, the orbital period, P, and the mid-eclipse time of reference, $T_0$, are the previously mentioned fitting parameters. $E_i$ denotes the epochs with respect to the mid-eclipse time of reference. Both orbital period and mid-eclipse time of reference determined in this work are listed in Table~\ref{tab:ephemeris} for the 26 eclipsing binary systems that were followed by TESS during the first year of observations to fulfill its timing verification. For the fitted parameters, errors are obtained from the 68.27\% confidence level of the marginalized posterior distribution. While the individual mid-eclipse times and their uncertainties are computed from the posterior distributions obtained from 10$^5$ MCMC steps, the ephemerides refinement are created by 10$^6$ MCMC steps. In both cases, we apply a conservative 25\% burn-in of the initial chains. For each of the binary systems we visually inspect the posterior distributions for normality. We check for convergence of the chains by sub-dividing the remaining 75\% in three, computing the usual statistics in each case, and checking for 1-$\sigma$ consistency in the periods and mid-eclipse times of reference. We carried out this procedure to reject the stars showing either a large spread in their O-C diagrams or intrinsic timing variability. Figure~\ref{fig:OC_TESS} shows, in red points, the O-C values of the binary systems that did not show a large spread. As usual, the O-C points were constructed subtracting to each mid-eclipse time (O) the mid-eclipse time assuming a constant period (C). As timing requirement, we considered a standard deviation of the O-C points smaller than 30 seconds. This limit rejected a few binary systems which O-C points were clearly showing intrinsic variability. The figure includes 405 O-C points, and has been made from primary eclipses only, as the secondary eclipses in most cases were shallow ($\Delta$Flux $\sim$ 0.1\%) and thus not providing timings as precise as their primary counterparts.

If a time drift is taking place in TESS photometry, this should be manifested equally in all the O-C points. In consequence, rather than fitting to the individual mid-eclipse times Eq.~\ref{eq:nosigma}, we considered the following expression, which was fitted to the 405 eclipse times jointly:

\begin{equation}
  T_{{\rm obs},i,j} = P_j E_{i,j} + T_{0,j} + \sigma_{\rm drift} \times (t-t_{\rm ref})\,,
  \label{eq:sigma}
\end{equation}
\noindent where $j$ runs over the different eclipsing binary systems, and $i$ over the number of eclipses for a given system. The shared constant drift amongst all systems is given by $\sigma_\mathrm{drift}$ in seconds per day, and $t_{\rm ref}$ is a common reference time for the drift. For a given system $P_j$, $T_{0,j}$ and $E_{i,j}$ correspond to the orbital period, time of reference, and eclipse ephemeris. In this work and for simplicity, we have always considered the drift to vary linearly with time, and we tested $\sigma_\mathrm{drift}$ against both a monotonic growth and decay. As a simple example to stress the power of the method, if we consider a time drift of \mbox{$\sigma_\mathrm{drift}$ = +1 second/day}, after a year of observations the last eclipse of a star located in the continuous viewing zone (CVZ) would be shifted about 6 minutes with respect to its non-shifted counterpart. This difference can easily be detected by eye when comparing contemporaneous observations from space and from the ground. However, a shift like this could pass unnoticed if only a single space-based data set is analyzed, rather than the whole sample. If, for instance, the individual mid-eclipse times of this time-drifted binary system are fitted only, the orbital period would slightly change when compared to that of a non-shifted data set, compensating for the drift. In this case, the individual O-C diagram would appear most likely flat, as only drifts of several seconds per day would create a curvature in an O-C diagram. In consequence, it is fundamental to carry out this exercise analyzing all the stars in the sample at once, because in doing so a small time drift could resurface from the noise.

\begin{figure}[ht!]
    \includegraphics[width=.5\textwidth]{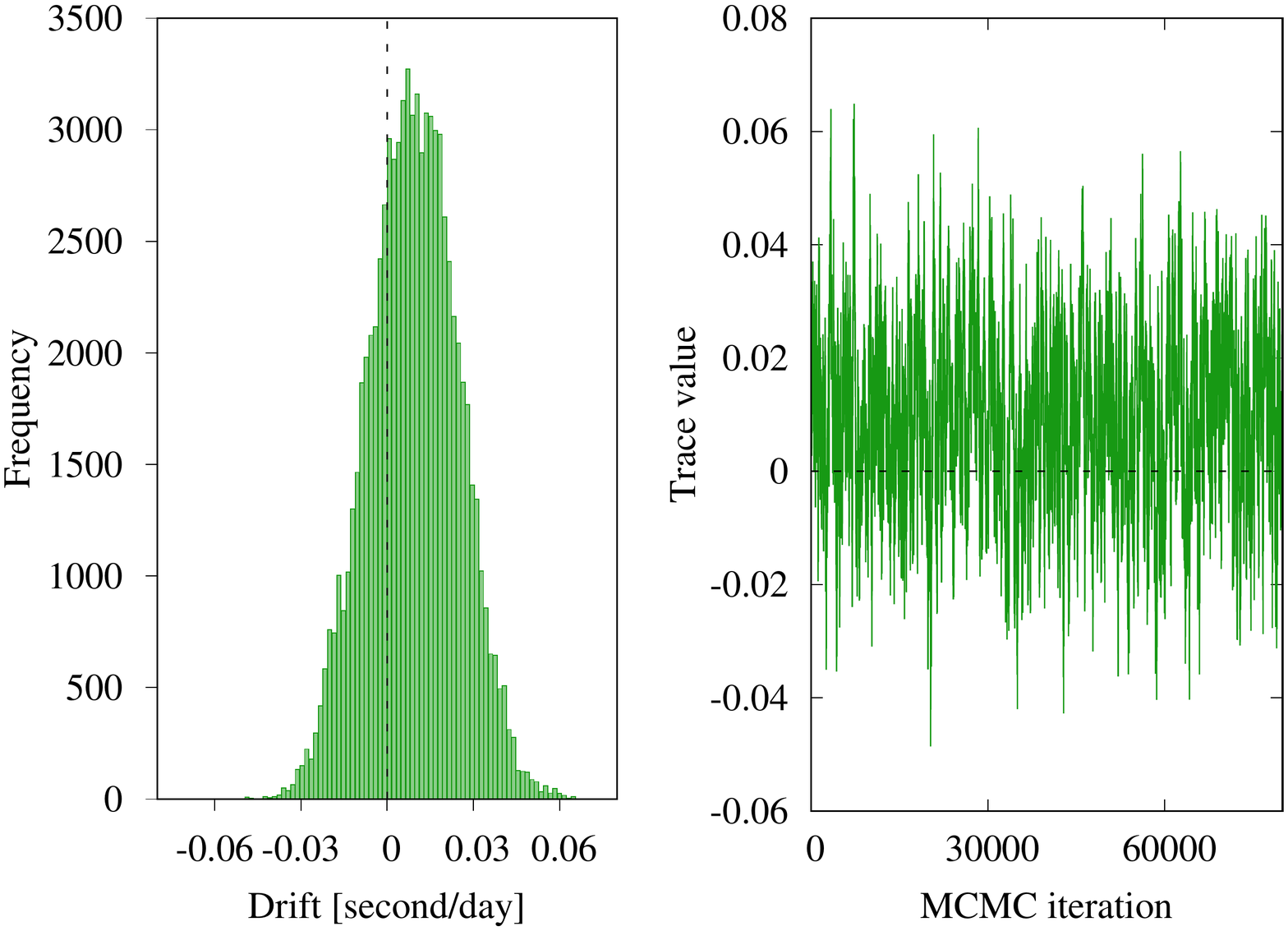}
    \caption{\label{fig:tracesigma} The resulting Markov chain for $\sigma_{\rm drift}$. Left: posterior distribution. Right: evolution of the traces.}
\end{figure}

Assuming then that there is a time drift growing (or decaying) linearly in time, we re-fitted the individual mid-eclipse times of all the stars together, but this time using Equation~\ref{eq:sigma} as model and, as previously mentioned, considering $\sigma_\mathrm{drift}$ to be equal for all the stars. As starting values for the period and the mid-eclipse times of reference we used the ones obtained before, assuming no shift, along with uniform priors covering $\pm50\%$ the starting values. If a time drift exists in TESS data, this would reflect into a $\sigma_\mathrm{drift}$ inconsistent with zero. After performing $10^6$ MCMC iterations along with a conservative burn-in of the first $20\%$ of the samples, the derived drift computed from the posterior distribution of the parameter was determined to be $\rm \sigma_\mathrm{drift} = 0.009 \pm 0.015$ seconds/day, fully consistent with zero at 1-$\sigma$ level. The corresponding posterior distribution and evolution of the traces can be seen in Figure~\ref{fig:tracesigma}. To allow for visual inspection, the triangle plots of twelve randomly selected best-fit individual periods and corresponding mid-eclipse times of reference can be seen in the Appendix, under Figures~\ref{fig:TP_per} and \ref{fig:TP_T0}, respectively, showcasing the rather low correlations between the parameters. 

\begin{figure}[ht!]
    \centering
    \includegraphics[width=.5\textwidth]{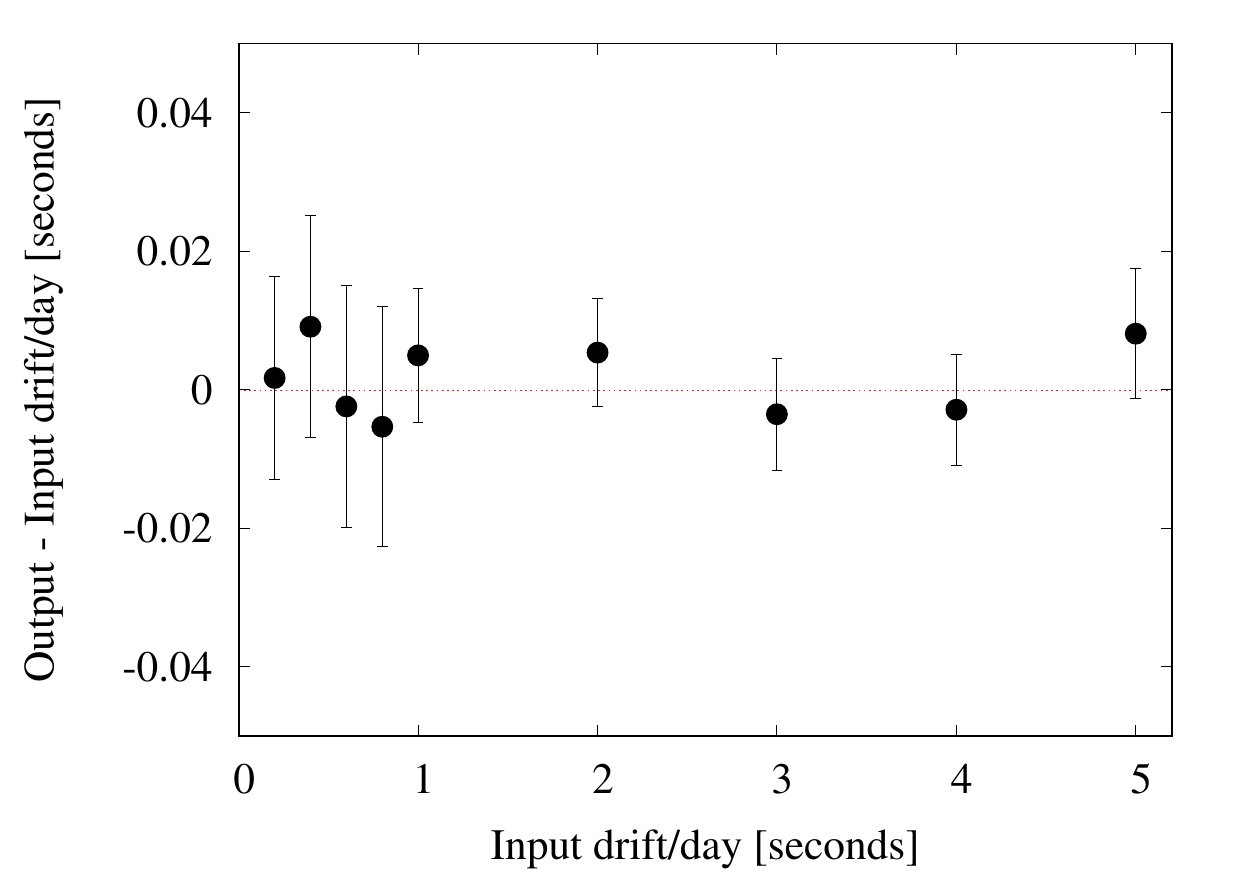}
    \caption{\label{fig:sigma_in_out}Injected versus recovered time drift. The figure shows the difference between the recovered and the injected drift, as a function of the injected drift. Uncertainties are given at 1-$\sigma$ level.}
\end{figure}

In order to test the robustness of our method, we injected into the TESS photometry drifts corresponding to 0.2, 0.4, 0.6, 0.8, 1, 2, 3, 4, and 5 seconds per day. If the technique works, we should be able to recover the injected signal with a similar precision as before. After shifting all TESS timestamps we re-computed the individual mid-eclipse times, we fitted the individual periods and mid-eclipse times of reference as if these shifts would not exist, and then we attempted at recovering the injected drift by fitting all the individual mid-eclipse times following Eq.~\ref{eq:sigma}. \fref{fig:sigma_in_out} shows the difference between the recovered and the injected time drifts, in seconds per day. As the figure shows, within uncertainties all the recovered values are fully consistent with the injected ones. 

\section{Conclusion}
\label{sec:DyC}

\begin{figure*}[ht!]
    \centering
    \includegraphics[width=\textwidth]{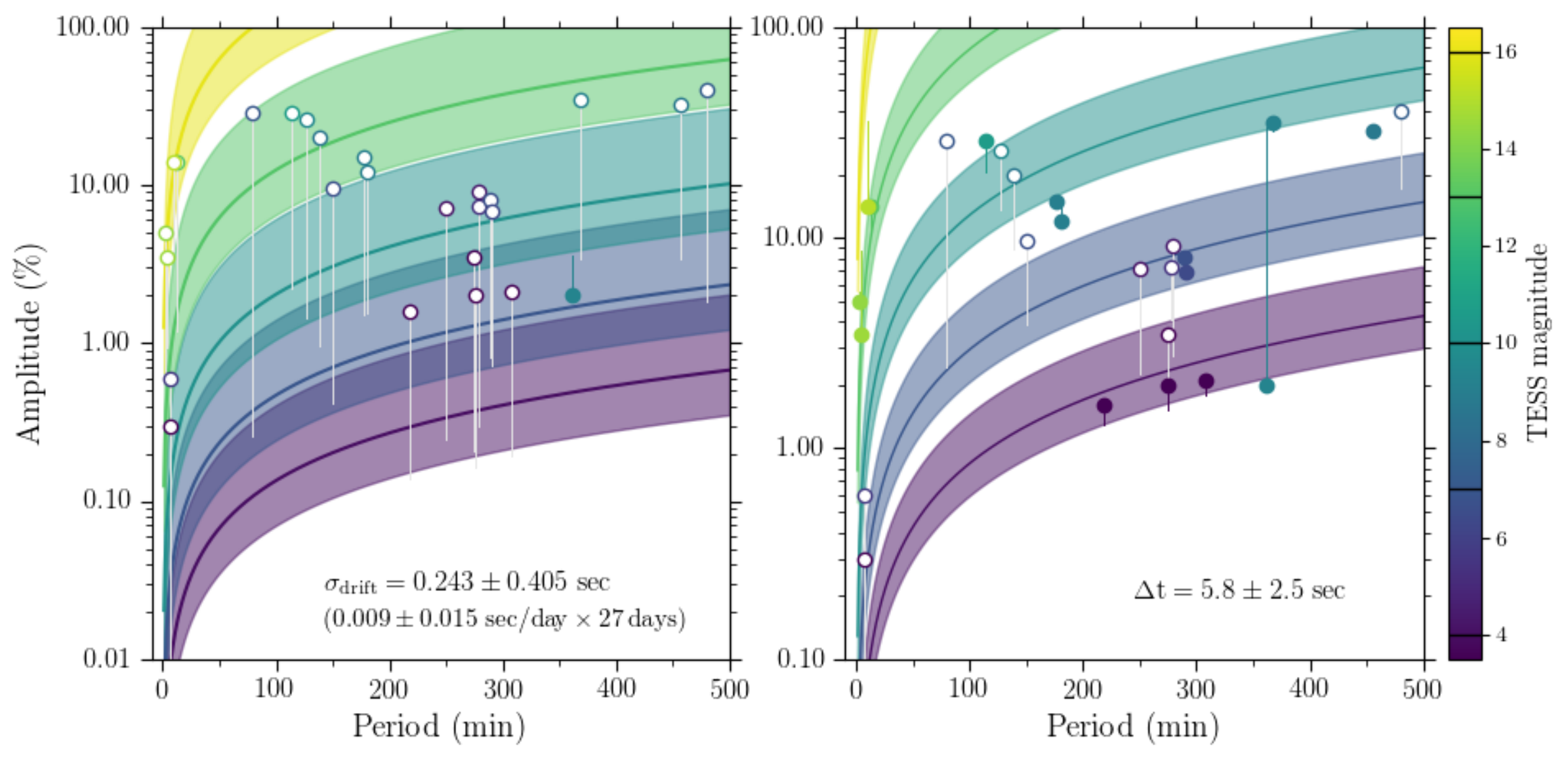}
    \caption{\label{fig:timelim} Reachable amplitudes (in relative variability in percent) as a function of oscillation period in minutes and TESS magnitude (colour). Markers indicate stars listed in the document SAC\_TESS\_0002\_5, for which a TESS magnitude could be found -- in cases where the measured timing values are of sufficient quality for asteroseismic analysis the marker is filled, with a colour corresponding to the TESS magnitude (i.e., where the stellar amplitude lie below the upper uncertainty limit of the computed relations for the given stellar magnitude and period), otherwise it is left open. Left: amplitudes corresponding to the measured time drift over a 27 day observing sector. The TESS magnitudes of the five relations shows are indicated on the colourbar. In each case the thick line gives the median value for the amplitude from a Monte Carlo sampling of the time drift (or offset) and the shaded region gives the corresponding uncertainty from the 16th and 84th percentiles. The vertical lines indicate for each star the median amplitude of the corresponding relation at the period and TESS magnitude of the star. Right: amplitudes corresponding to the measured absolute time offset.}
\end{figure*}

To reliably carry out asteroseismology studies from TESS data, potential drifts and the absolute time for TESS observations must be known to a high accuracy. Even though the internal clock on board TESS is very accurate in its own time, as it happened to the {\it Kepler} space telescope drifts and offsets could take place. In consequence, we have carried out a photometric follow-up of several eclipsing binary systems from TESS and from the ground, using two telescopes located at the Complejo Astron\'omico El Leoncito, in Argentina. Comparing the timings of twelve primary eclipses of binary systems of Algol type from the ground to those observed by TESS we find a time offset of 5.8 $\pm$ 2.5 seconds (in the sense that the barycentric time measured by TESS is ahead of real time), indicative of a small offset but still consistent with zero at the $2.3-\sigma$ level. It is worth to mention that the TESS team has recently discovered a time offset of 2 seconds that accounts for some portion of our detected time offset. As of sector 20, the data products on MAST are corrected. Taking this offset into consideration improves our results to a total time offset of 3.8 $\pm$ 2.5 seconds, consistent with zero at the $1.5-\sigma$ level. Carrying out a joint analysis of 405 individual mid-eclipse times collected from 26 eclipsing binary systems, we find TESS to have a time drift consistent with zero, and equal to $\sigma_{\rm drift} = 0.009 \pm 0.015$ seconds/day. For this, we assumed a monotonic, linearly growing --and decaying-- time-dependent drift. To the precision that our joined data can achieve, we can confirm that the TESS clock does not present neither a clear time offset nor a time drift. 

It is clear that we cannot reach a precision on the estimation of the time drift or offset that satisfy the requirements given in \sref{sec:req}. It is, however, worth remembering that these were defined based on the very brightest, highest amplitude, and shortest period pulsators. So, while our current analysis cannot guarantee TESS observations with timing specifications that ensure an optimum asteroseismic analysis for these, there will still be many fainter, lower amplitude, longer period pulsators whose requirements are fulfilled. In \fref{fig:timelim} we show the amplitudes that can be reached for a given pulsation period and TESS magnitude given the estimated drift and absolute offset. Given the relatively large uncertainties on our estimates the amplitude values were obtained from a Monte Carlo sampling rather than using standard error propagation. To compute the noise per measurement that enters in the calculations we used the prescription by \citet{Sullivan2015}, even though we are aware that the mission will do better than the estimates here. We combined this with measured values from the TASOC pipeline for mean flux and number of pixels in an aperture as a function of TESS magnitude \citep{TASOC}. We adopt a systematic noise of $5 \rm \, ppm\, hr^{-1}$, which mainly affect the noise at the very bright end ($T_{\rm mag} \lesssim 4$).

As seen from \fref{fig:timelim} it will be possible to compare stellar oscillations observed by TESS with ground-based observations for several of the stars listed in SAC\_TESS\_0002\_5 based on the measured absolute time offset. We find that the measured offset is of a size that will not become an issue for comparing ground-based and space data for coherent oscillations for most of the targets observed with TESS. Specifically we find that for all TESS stars fainter than $T_{\rm mag} =4$ oscillations with periods longer than one hour and amplitudes below ${\sim}5$ mmag ($0.5 \%$) are unaffected. For stars fainter than $T_{\rm mag} =9$ oscillations with periods longer than on hour and amplitudes below ${\sim}50$ mmag ($5 \%$) are unaffected.

Only for one of the stars in SAC\_TESS\_0002\_5 does the measured time drift allow for the theoretical accuracy to be reached. In the case of solar-like oscillators, with amplitudes of a few ppm and periods of the order of a few minutes on the main-sequence, to a few hundred ppm and periods of the order a day on the red-giant branch, the current timing measurements are sufficient to reach the theoretical accuracy on the determination of frequencies and comparison with ground-based facilities.

We note that the pulsators listed in SAC\_TESS\_0002\_5 represent some of the stars with the very strongest timing requirements within their respective variability class, and the requirements for most stars observed by TESS will therefore be less strict. Also, the model used for the photometric noise represents the lower envelope, so for many stars the photometry will be noisier and as a consequence the timing requirement will be reduced.

\acknowledgments
{
Funding for the Stellar Astrophysics Centre is provided by The Danish National Research Foundation (Grant DNRF106). The work is based on data obtained at Complejo Astron\'omico El Leoncito, operated under agreement between the Consejo Nacional de Investigaciones Cient\'ificas y T\'ecnicas de la Rep\'ublica Argentina and the National Universities of La Plata, C\'ordoba and San Juan. CvE and HK acknowledge support from the European Social Fund via the Lithuanian Science Council (LMTLT) grant No. 09.3.3-LMT-K-712-01-0103. MNL and RH acknowledge support from the ESA PRODEX programme.}

\bibliography{vonEssenC}
\bibliographystyle{aasjournal}

\appendix

\facilities{CASLEO:HSHT, CASLEO:JST}

\software{This work made use of \texttt{PyAstronomy}\footnoteref{note1}, \texttt{PyMC} \citep{Patil2010}, \texttt{SciPy} \citep{Jones2001}, matplotlib \citep{matplotlib}, numpy, \texttt{IRAF} \citep{iraf1993}, DIP$^2$OL \citep{vonEssen2018}}

\begin{table*}[ht!]
    \centering
    \caption{\label{tab:ephemeris} Orbital period ($P$) and mid-eclipse time ($T_0$ of reference for the 25 binary systems analyzed in this work. Uncertainties are given at 1-$\sigma$ level.}
    \begin{tabular}{l c c}
    \hline\hline
    Name            &                 $P$                    &              $T_0$                   \\
                    &               (days)                 &          BJD$_\mathrm{TDB}$ -2457000 \\
    \hline
    
    DW Aps & 2.312969 $\pm$ 3.3$\times$10$^{-6}$ & 1626.66503 $\pm$ 0.00004  \\
    V379 Cen & 1.874688 $\pm$ 1.5$\times$10$^{-5}$ & 1599.68944 $\pm$ 0.00009  \\
    WY Cet & 1.939802 $\pm$ 1.8$\times$10$^{-5}$ & 1387.98849 $\pm$ 0.00010    \\
    TZ Eri & 2.606213 $\pm$ 1.9$\times$10$^{-5}$ & 1439.30912 $\pm$ 0.000125   \\
    SU For & 2.434594 $\pm$ 3.5$\times$10$^{-5}$ & 1386.75724 $\pm$ 0.00013    \\
    RX Hya & 2.281730 $\pm$ 2.7$\times$10$^{-5}$ & 1518.10474 $\pm$ 0.00011    \\
    RR Nor & 1.5137439 $\pm$ 2.1$\times$10$^{-6}$ & 1625.17032 $\pm$ 0.00003   \\
    GT Vel & 4.6700996 $\pm$ 9.2$\times$10$^{-6}$ & 1520.12020 $\pm$ 0.00004   \\
    UW Vir & 1.810798 $\pm$ 5.6$\times$10$^{-5}$ & 1572.62214 $\pm$ 0.00036    \\
    UY Vir & 1.9943626 $\pm$ 5.1$\times$10$^{-6}$ & 1571.16659 $\pm$ 0.00003   \\
    V636 Cen & 4.283994 $\pm$ 7.0$\times$10$^{-5}$ & 1598.95471 $\pm$ 0.00016  \\
    V646 Cen & 2.246539 $\pm$ 3.6$\times$10$^{-5}$ & 1572.84139 $\pm$ 0.00027  \\
    AF Cru & 1.895661 $\pm$ 1.2$\times$10$^{-5}$ & 1599.39700 $\pm$ 0.00008    \\
    OU Lup & 4.610498 $\pm$ 6.9$\times$10$^{-5}$ & 1601.87579 $\pm$ 0.00011    \\
    BH Pup & 1.915908 $\pm$ 7.5$\times$10$^{-5}$ & 1519.08404 $\pm$ 0.00041    \\
    TV Nor & 8.52456 $\pm$ 0.00014               & 1625.60512 $\pm$ 0.00017    \\
    YZ Ant & 2.152446 $\pm$ 2.7$\times$10$^{-5}$ & 1546.45178 $\pm$ 0.00014    \\
    BV Ant & 3.594289 $\pm$ 1.8$\times$10$^{-5}$ & 1546.43852 $\pm$ 0.00012    \\
    BD Dor & 0.78524198 $\pm$ 3.8$\times$10$^{-7}$ & 1545.56357 $\pm$ 0.00003  \\
    AT Men & 2.3446214 $\pm$ 1.0$\times$10$^{-6}$ & 1411.55316 $\pm$ 0.00006   \\
    DE Phe & 1.4029532 $\pm$ 2.2$\times$10$^{-6}$ & 1354.37999 $\pm$ 0.00001   \\
    X Pic & 0.86189657 $\pm$ 1.2$\times$10$^{-7}$ & 1386.63236 $\pm$ 0.00002   \\
    AO Pic & 2.23418239 $\pm$ 3.9$\times$10$^{-7}$ & 1327.56424 $\pm$ 0.00003  \\
    FU Vel & 2.446837 $\pm$ 1.8$\times$10$^{-5}$ & 1545.43213 $\pm$ 0.00019    \\
    EQ Vel & 1.0802739 $\pm$ 2.1$\times$10$^{-6}$ & 1517.76048 $\pm$ 0.00005   \\
    NV Tel & 3.545012 $\pm$ 8.9$\times$10$^{-5}$ & 1659.73105 $\pm$ 0.00026    \\
    \hline
    \end{tabular}
\end{table*}

\begin{figure*}[ht!]
    \centering
    \includegraphics[width=\textwidth]{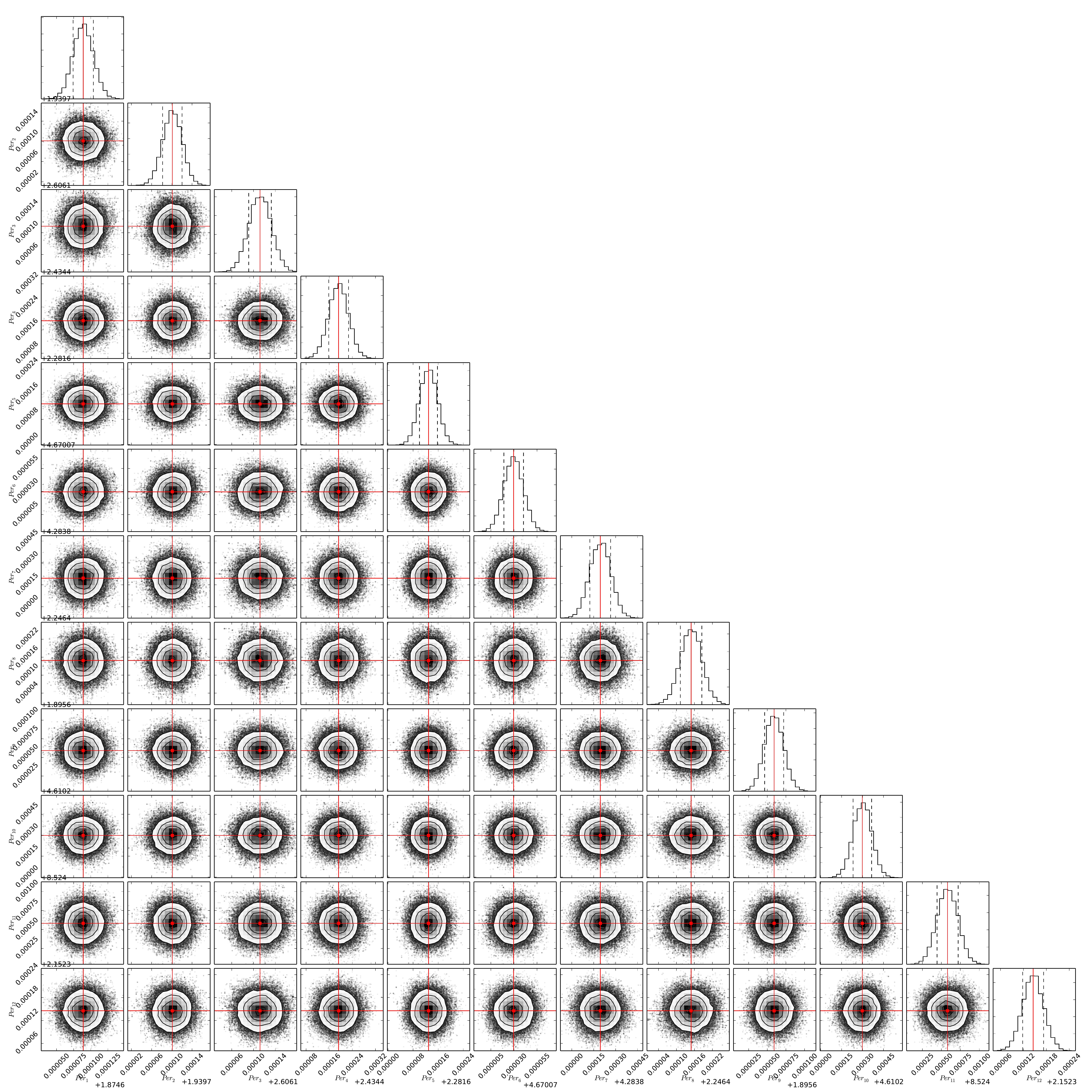}
    \caption{\label{fig:TP_per} A triangle plot for twelve randomly selected orbital periods of eclipsing binaries observed by TESS. Red points correspond to the best-fit parameters and shaded gray to white areas correspond to 1, 2, and 3-$\sigma$ uncertainty regions.}
\end{figure*}

\begin{figure*}[ht!]
    \centering
    \includegraphics[width=\textwidth]{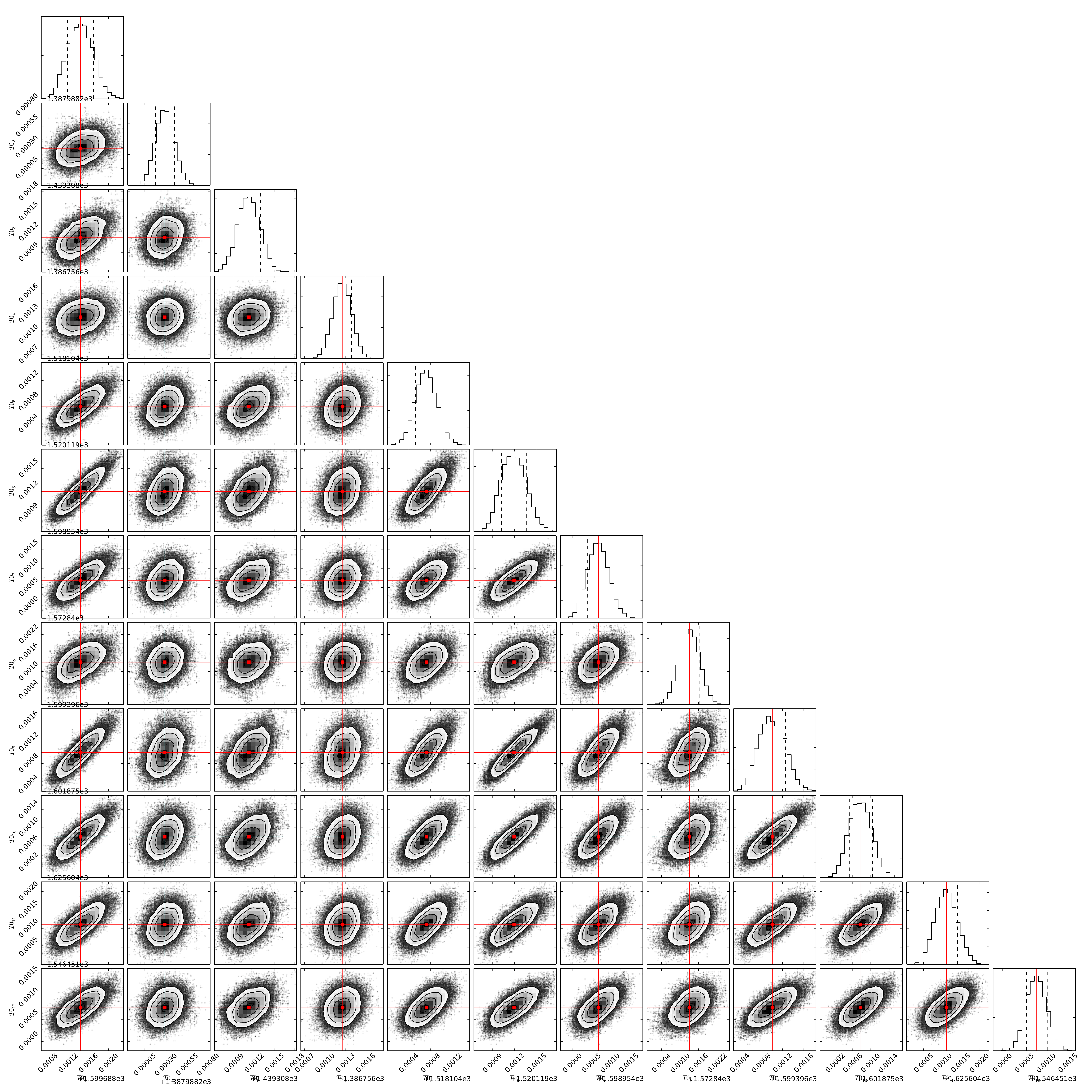}
    \caption{\label{fig:TP_T0}Same as Fig.~\ref{fig:TP_per}, but for the corresponding mid-eclipse times of reference.}
\end{figure*}

\end{document}